\documentclass[twocolumn]{aastex61}
\usepackage{amsmath}
\usepackage{subfigure}
\graphicspath{{Img/},{./}}

\begin{document}
\title{Dissection of the collisional and collisionless mass components in a mini sample of CLASH and HFF massive galaxy clusters at $z \approx 0.4$}

\author{M.~Bonamigo}
\affiliation{Dark Cosmology Centre, Niels Bohr Institute, University of Copenhagen, Juliane Maries Vej 30, DK-2100 Copenhagen, Denmark} 

\author{C.~Grillo}
\affiliation{Dipartimento di Fisica, Universit\`a degli Studi di Milano, via Celoria 16, I-20133 Milano, Italy} 
\affiliation{Dark Cosmology Centre, Niels Bohr Institute, University of Copenhagen, Juliane Maries Vej 30, DK-2100 Copenhagen, Denmark} 

\author{S.~Ettori}
\affiliation{INAF, Osservatorio Astronomico di Bologna, Via Piero Gobetti, 93/3, 40129 Bologna, Italy}
\affiliation{INFN, Sezione di Bologna, viale Berti Pichat 6/2, 40127 Bologna, Italy}

\author{G.~B.~Caminha}
\affiliation{Kapteyn Astronomical Institute, University of Groningen, P.O. Box 800, 9700 AV, Groningen, The Netherlands}
\affiliation{Dipartimento di Fisica e Scienze della Terra, Universit\`a degli Studi di Ferrara, Via Saragat 1, I-44122 Ferrara, Italy} 

\author{P.~Rosati}
\affiliation{Dipartimento di Fisica e Scienze della Terra, Universit\`a degli Studi di Ferrara, Via Saragat 1, I-44122 Ferrara, Italy} 

\author{A.~Mercurio}
\affiliation{INAF - Osservatorio Astronomico di Capodimonte, Via Moiariello 16, I-80131 Napoli, Italy} 

\author{E.~Munari}
\affiliation{INAF - Osservatorio Astronomico di Trieste,  Via G.B. Tiepolo, 11 I-34143 Trieste, Italy} 

\author{M.~Annunziatella}
\affiliation{Department of Physics and Astronomy, Tufts University, 574 Boston Avenue, Medford, MA 02155, USA}
\affiliation{INAF - Osservatorio Astronomico di Trieste,  Via G.B. Tiepolo, 11 I-34143 Trieste, Italy} 

\author{I.~Balestra}
\affiliation{University Observatory Munich, Scheinerstrasse 1, D-81679 Munich, Germany}

\author{M.~Lombardi}
\affiliation{Dipartimento di Fisica, Universit\`a degli Studi di Milano, via Celoria 16, I-20133 Milano, Italy}

\defcitealias{PlanckXIII}{Planck collaboration XIII 2016}
\defcitealias{Bonamigo2017}{BO17}

\email{bonamigo@dark-cosmology.dk}

\begin{abstract}
We present a multi-wavelength study of the massive ($M_{200\textrm{c}} \approx 1$-$2 \times 10^{15} M_\odot$)
galaxy clusters RXC~J2248.7$-$4431, MACS~J0416.1$-$2403, and MACS~J1206.2$-$0847
at $z \approx 0.4$.
Using the X-ray surface brightness of the clusters from deep Chandra data to
model their hot gas, we are able to disentangle this mass term from the diffuse
dark matter in our new strong-lensing analysis, with approximately $50$-$100$
secure multiple images per cluster, effectively separating the
collisional and collisionless mass components of the clusters.
At a radial distance of $10\%$ of $R_{200\textrm{c}}$ (approximately $200$ kpc), we
measure a projected total mass of $(0.129 \pm 0.001)$, $(0.131 \pm 0.001)$ and
$(0.137 \pm 0.001)\times M_{200\textrm{c}}$, for RXC~J2248, MACS~J0416 and MACS~J1206,
respectively.
These values are surprisingly similar, considering the large differences in the
merging configurations, and, as a consequence, in the mass models of the clusters.
Interestingly, at the same radii, the hot gas over total mass fractions differ substantially, ranging
from $0.082 \pm 0.001$ to $0.133 \pm 0.001$, reflecting the various dynamical
states of the clusters.
Moreover, we do not find a statistically significant offset between the
positions of the peak of the diffuse dark matter component and of the BCG in the more complex clusters
of the sample.
We extend to this sample of clusters previous findings of a number of massive
sub-halos higher than in numerical simulations.
These results highlight the importance of a proper separation of the
different mass components to study in detail the properties of dark matter in galaxy
clusters.
\end{abstract}


\section{Introduction} \label{sec:intro}
In recent years, many observational campaigns have targeted massive galaxy
clusters to study and model their gravitational lensing of background sources.
While some surveys, like the Cluster Lensing And Supernova survey with Hubble
\citep[CLASH;][]{Postman2012}, have focused on the investigation of the dark
matter (DM) halos in which clusters live, others, like the Hubble Frontier Fields \citep[HFF;][]{Lotz2017}
and the Reionization Lensing Cluster Survey \citep[RELICS;][]{Salmon2017},
have turned clusters into powerful gravitational telescopes to explore the
high-$z$ Universe.
New imminent surveys (i.e., BUFFALO; P.I. Steinhardt, C.) are planning to expand this field even further and
several GTO programs of the coming \emph{James Webb Space Telescope} have already been
scheduled to the study of lensing galaxy clusters.
Ground-based photometric and spectroscopic data have been used to complement
the space-based observations.
For instance, within the CLASH-VLT program \citep{Rosati2014}, thousands of
member galaxies and lensed multiple images have been spectroscopically confirmed
\citep[e.g.,][]{Biviano2013,Balestra2016,Monna2017}.
In the cores of clusters, the Multi Unit Spectroscopic Explorer
\citep[MUSE;][]{Bacon2012} at the VLT has also allowed for the serendipitous discovery
of lensed systems that are not detected in the HST images
\citep{Richard2015,Caminha2017b,Karman2017,Mahler2018}, in addition to the redshift measurement of
previously known objects.
Moreover, the same massive galaxy clusters have also been the targets of numerous observations with X-ray
telescopes, like \emph{Chandra} and \emph{XMM Newton}, and submillimeter and
radio antennas, that have characterized with extreme precision the hot
intracluster gas component \citep[e.g.,][]{Donahue2014,Ogrean2016,Rumsey2016,vanWeeren2017}.

All these studies have helped to create a multi-wavelength view of galaxy
clusters.
Multi-probe analyses have the advantage of highlighting systematic errors 
\citep[or lack thereof; see e.g.,][]{Balestra2013}, breaking degeneracies \citep[i.e.\ line-of-sight projection; see e.g.,][]{Morandi2012,Umetsu2015,Sereno2017},
and separating the cluster components \citep[i.e., diffuse DM, hot intracluster
gas and cluster member galaxies; see e.g.,][]{Annunziatella2017}.
The latter is the aim of this paper.
Following \citet{Bonamigo2017}, we combine information from gravitational
lensing and X-ray observations, in order to isolate the different mass
components in a small sample of galaxy clusters: RXC~J2248.7$-$4431,
MACS~J0416.1$-$2403, and MACS~J1206.2$-$0847.
This mass dissection allows for a more direct comparison with numerical
simulations and theoretical predictions, as the reconstructed mass components,
with different characteristics, i.e. different physical properties and spatial
distributions, are less contaminated with each other.
An unbiased characterization of the cluster components can shed light on the
nature of DM \citep[i.e., the Bullet cluster, 1E 0657$-$56;][]{Clowe2006}
and be used to measure the values of the cosmological parameters.
For example, the shape of the mass density profile is related to the cluster
formation history and the value of the DM cross-section, through the density profile
characteristic radius \citep{Wechsler2002,Zhao2003} and inner slope
\citep{Spergel2000,Firmani2000,Vogelsberger2012,Maccio2012}.
Similarly, self-interacting DM can produce an offset between the center of
its mass distribution and the truly-collisionless galaxy distribution in a cluster.
Moreover, the fraction of baryons in galaxy clusters can be used to infer the
background value, the cosmological baryon fraction $\Omega_b$
\citep{White1993,Evrard1997,Ettori2003,Allen2008,Planelles2013}.

The paper is organized as follows.
In Section \ref{sec:method}, we briefly summarize the technique used to separate
the cluster mass components in our analysis.
Then, in Section \ref{sec:sample}, we introduce the studied sample of galaxy clusters.
Section \ref{sec:results} contains the results of the X-ray surface brightness
and strong-lensing analyses, on an individual cluster basis; while,
in Section \ref{sec:discussion}, we compare the results across the whole sample.
Finally, in Section \ref{sec:conclusions}, we summarize our conclusions.

Throughout the paper, we adopt a flat $\Lambda$CDM cosmology with Hubble
constant $H_0 = 70$ km s$^{-1}$ Mpc$^{-1}$ and total matter density
$\Omega_m = 0.3$.
All magnitudes are given in the AB system.

\section{Method}\label{sec:method}
The method we adopt to separate the different cluster mass components consists of two
steps: first, we fit the X-ray surface brightness to obtain a mass model of the hot
intracluster gas, then, we include this in the strong-lensing analysis
as a fixed component, to derive the DM and member galaxy mass distributions.

In the following subsections, we will briefly summarize these two steps; a
more detailed presentation of the method can be found in \citet{Bonamigo2017}.

\subsection{X-ray}
We choose to describe the mass distribution of the hot intracluster gas
with a combination of several dual Pseudo Isothermal Ellipsoidal \citep[dPIE]{Eliasdottir2007,Suyu2010}
distributions, as this profile is already available in most gravitational lensing
softwares.
In order to convert the gas mass density into its corresponding X-ray surface brightness, we compute
the cooling function, which gives the number of emitted photons per unit of volume.
In the energy range considered in this work ($0.7-2$~keV), the dependence of the
cooling function on the values of the gas temperature and metallicity is very weak
\citep{Ettori2000}, therefore, we can safely adopt constant values for these
quantities throughout the whole observed region of a cluster.
Within these assumptions, the X-ray surface brightness is proportional, via
the cooling function, to the the squared mass density projected along the
line of sight.

From the median values of the gas temperature and metallicity, derived from the measured
radial profiles, we compute the cooling function, using the Astrophysical Plasma
Emission Code (APEC\footnote{\url{http://atomdb.org/}}) model.
Then, we measure the background emission from an image region at a projected
distance of more than $1.5$ Mpc from the cluster center and use a photoelectric absorption
(phabs\footnote{\href{http://heasarc.gsfc.nasa.gov/docs/xanadu/xspec/manual/XSmodelPhabs.html}{Xspec manual: phabs}})
model for the foreground galactic gas. 
Finally, we fit the Chandra image, reduced and corrected for exposure, using the
software \emph{Sherpa}\footnote{\url{http://cxc.harvard.edu/sherpa}}.

\subsection{Strong lensing}\label{sec:lensing}
After creating a model for the hot intracluster gas mass, we proceed to include it
in the cluster strong-lensing analysis.
As detailed in \citet{Bonamigo2017}, we have included the hot gas term
as a fixed component. This is justified by the small set of assumptions required
to derive the hot gas mass density profiles from the X-ray surface brightness
and by the fact that the statistical errors on the inferred hot gas mass density
profiles are smaller than those typically associated with the other cluster mass
components.
The model of the cluster total mass consists of three components: large-scale
DM halos, the hot intracluster gas, and galaxy-scale halos.
The first term describes the diffuse DM that spans the whole cluster and accounts for
most of its total mass.
For this component we use one or more Pseudo-Isothermal Elliptical Mass
Distribution \citep[hereafter PIEMD;][]{Kassiola1993} profiles, the
number of the profiles depending on the degree of complexity of the cluster mass
distribution.
Each PIEMD profile has $6$ free parameters: center position, $x_h$ and
$y_h$, ellipticity, $\epsilon_h$, position angle, $\theta_h$, core radius, $R_{C,h}$,
and central velocity dispersion, $\sigma_{0,h}$.
The galaxy-scale halos are modeled instead with spherical dPIE distributions,
with their centers fixed on the luminosity centers of the cluster galaxies, thus
resulting in two free parameters for each galaxy: truncation radius $R_{T,i}$
and central velocity dispersion $\sigma_{0,i}$.
To reduce the otherwise too large number of free parameters, we scale the values of  
$R_{T,i}$ and $\sigma_{0,i}$ depending on $L_i$, i.e., the galaxy luminosity in the
HST/WFC3 filter F160W, with respect to a reference luminosity, $L_g$, so that
$R_{T,i} = R_{T,g} ( L_i/L_g )^{0.5}$ and
$\sigma_{0,i} = \sigma_{0,g} ( L_i/L_g )^{0.35}$.
With these scaling relations, which reproduce the tilt of the fundamental plane
of elliptical galaxies \citep{Faber1987,Bender1992}, all the galaxy-scale halos
are parametrized by only two quantities: $\sigma_{0,g}$ and $R_{T,g}$.

To infer the values of all the model parameters of the mass components of a
cluster, we use the lensing software \emph{lenstool} \citep{Jullo2007}.
We run an initial optimization on the positions of several tens of
multiple images with positional errors of $0\arcsec.5$ and $1\arcsec$,
respectively, for images detected on the HST- or MUSE-only data.
Then, we multiply these errors by a constant factor, obtained by requiring
that the best-fit $\chi^2$ value gets approximately equal to the number of
degrees of freedom (d.o.f.) of the model (see Appendix
\ref{sec:chi2distribution} for further quantitative discussion on the
implications of this assumption).
We use the updated values of the positional errors to finally
sample the posterior distribution of the cluster model parameters.
By doing so, we make sure not to under- or over-estimate the uncertainties on
the model parameters and to include possible systematic effects,
such as line-of-sight mass components or unresolved substructures.
A recent work \citep{Acebron2017} has shown that an underestimated value for the
positional error can lead to biased analyses and that the bias decreases when
the $\chi^2$ value is close to the number of d.o.f.

\section{The sample}\label{sec:sample}
The sample studied in this work is composed of three galaxy clusters:
RXC J2248.7$-$4431 (also known as Abell S1063), MACS J0416.1$-$2403, and MACS
J1206.2$-$0847.
Hereafter, we will use the shortened names RXC~J2248, MACS~J0416, and MACS~J1206,
respectively.
For these objects some of the best state-of-the-art observations are available:
multi-band HST imaging, VLT/MUSE and VIMOS spectroscopic data, and deep Chandra
observations.
Indeed, thanks to the CLASH-VLT program \citep{Rosati2014}, spectra for a large
number of sources are available, on the order of thousands per cluster field.
Moreover, all three targets have been observed for several hours with the
MUSE integral-field spectrograph: two pointings of 3.1 and 4.8 hours each in
RXC~J2248 (ID 060.A-9345(A) and 095.A-0653(A), P.I.: K. Caputi), two pointings of 2 and 11
hours in MACS~J0416 (ID 094.A-0115(B), P.I.: J. Richard, and ID 094.A-0525(A)
P.I.: F.E. Bauer), and three pointings of about 4 hours in MACS~J1206 (ID
095.A-0181(A) and 097.A-0269(A), P.I.: J. Richard).
Additionally, two of them, RXC~J2248 and MACS~J0416, are also part of the Hubble
Frontier Fields sample \citep{Lotz2017}.
Such datasets make these clusters the ideal candidates for 
accurate strong-lensing analyses \citep{Caminha2016a,Grillo2016,Caminha2017a,Caminha2017b,Lagattuta2017},
which, in turn, can be used as the foundation for the multiwavelength analysis
we perform in this work.

To build a precise model of the hot intracluster gas, we use deep Chandra
observations.
The combined exposure times are $123$~ks (obsID  4966, 18611 and 18818), $293$~ks 
(obsID 16236, 16237, 16304, 16523, 17313) and $23$~ks (obsID 3277), for
RXC~J2248, MACS~J0416 and MACS~J1206, respectively.
A detailed X-ray analysis of MACS0416 is given in \citet{Ogrean2016}.
All images are reduced with the software \emph{CIAO} (version 4.7+) and using
the calibration database \emph{CALDB} (version 4.6.8+).
The resolution of the surface brightness maps, limited to the energy range from
$0.7$~keV to $2$~keV, is scaled down to a pixel size of $1\arcsec.968$
($3\arcsec.936$ for MACSJ~0416).
This pixel size is much larger than \emph{Chandra}'s on-axis point-spread function;
therefore, we do not consider this effect in our analysis.

Accurate strong lensing models rely on highly complete and pure cluster-member
catalogs and samples of secure multiple images that cover a broad range of
redshifts.
By combining the information from the HST, VIMOS and MUSE data, we are
able to produce such catalogs.
In particular, we start from the spectroscopically-confirmed cluster members to define a
color-space region which we then use to derive the probability of a galaxy to
belong to the cluster \citep{Grillo2015}.
The resulting catalogs of cluster members have
a completeness value of approximately $95\%$ \citep{Grillo2015,Caminha2017b} and
are comprised mostly of spectroscopically confirmed members.
Conversely, the multiple-image catalogs consist only of secure
spectroscopically confirmed sources \citep{Balestra2016,Caminha2016a,Caminha2017b}.

The galaxy cluster RXC~J2248, first identified as Abell S1063 by \citet{Abell1989}, is the most massive,
$M_{200\textrm{c}}$\footnote{The mass $M_{200\textrm{c}}$ is defined as the mass within a sphere
inside which the value of the mean density is equal to $200$ times that of the
Universe critical density at each cluster redshift.} $ = (2.03 \pm 0.67) \times 10^{15} M_\odot$ \citep{Umetsu2014},
and the nearest in the sample; at its redshift, $0.348$, $1\arcsec$ corresponds
to $4.92$ kpc.
It is also the least complex, with an almost unimodal total mass
distribution and a single BCG (R.A. $=$ 22:48:43.970 and decl. $= -$44:31:51.16).
The X-ray surface brightness is quite symmetric, although it is slightly
offset from cluster center, and it shows a cool core, that instead coincides with
the position of the BCG.
In total, we use $55$ multiple images (all spectroscopically confirmed) from
$20$ background sources, covering a redshift range from $0.73$ to $6.11$
\citep[extending the sample by][]{Karman2017}.
We describe the total mass distribution of RXC~J2248 with a model, similar to
that presented in \citet{Caminha2016a}, which consists of a large-scale elliptical
PIEMD halo (DM and intracluster light), three elliptical dPIE components (hot
gas), $222$ galaxy-scale dPIE halos (member galaxies), and an additional
small-scale spherical halo.
The latter was initially centered on the location of a small group of galaxies,
but, in the final optimized model, its position does not coincide with any
particular feature of the cluster.
This additional component reduces the offset between the observed and
model-predicted positions of some multiple images in the North-East region and
it has been introduced also in the model of RXC~J2248 by \citet{Kawamata2017}.
It is assumed to have a spherical singular isothermal density
profile with $3$ free parameters: center position, $x_{h2}$ and $y_{h2}$, and
central velocity dispersion, $\sigma_{0,h2}$.
The reference galaxy for the cluster-member scaling relations is the BCG.

MACS~J0416 is a merging cluster, with a $M_{200\textrm{c}}$ mass value of approximately
$(1.04 \pm 0.22) \times 10^{15} M_\odot$ \citep{Umetsu2014}, and located at a redshift of
$0.396$, where $1\arcsec$ corresponds to $5.34$~kpc.
It was first discovered in the Massive Cluster Survey (MACS) by \citet{Mann2012}.
It hosts two BCGs, G1 and G2, located, respectively, in the northeast
(R.A. $=$ 04:16:09.154 and decl. $= -$24:04:02.90) and southwest
(R.A. $=$ 04:16:07.671 and decl. $= -$24:04:38.75) regions of the cluster.
Its merging status is evident from the X-ray emission morphology and the large
projected separation ($\sim 200$ kpc) between the two BCGs.
In the strong-lensing analysis, we use $102$ spectroscopically-confirmed
multiple images from $37$ background sources, with redshifts from $0.94$ to
$6.15$ \citep{Caminha2017a}. 
The mass model is an update of the model used in \citet{Bonamigo2017}
\citep[itself derived from][]{Grillo2015,Caminha2017a} and consists of three
large-scale PIEMD halos (DM and intracluster light), four elliptical dPIE
components (hot gas), and $193$ galaxy-scale dPIE halos (member galaxies,
including the two BCGs).
Two of the large-scale halos are elliptical in projection and describe the two
merging subclusters, while the third component traces the mass of a small group
of galaxies present in the North-East region of the cluster.
This halo is assumed to be spherical with $4$ free parameters: center position,
$x_{h3}$ and $y_{h3}$, core radius, $R_{C,h3}$, and central velocity dispersion,
$\sigma_{0,h3}$.
Moreover, we use the northern BCG, G1, as the reference galaxy for the
scaling relations that define the properties of the cluster member galaxies.
Finally, an additional galaxy-scale mass component takes into account the
lensing perturbation introduced by a foreground galaxy (R.A. $=$ 04:16:06.82 and
decl. $= -$24:05:08.4) at redshift $0.112$.
This galaxy is described by a dPIE profile at the redshift of the cluster and therefore
the values of $\sigma_0$ and $R_T$ should be considered only as effective
parameters.
As shown by \citet{Chirivi2017}, the introduction of this foreground galaxy at
the cluster redshift gives results that are very similar to a full multi-plane
analysis, both in terms of the inferred cluster parameter values and offset
between the observed and model-predicted positions of the multiple images.

Finally, the galaxy cluster MACS~J1206 has a $M_{200\textrm{c}}$ mass value of
approximately $(1.59 \pm 0.36) \times 10^{15} M_\odot$ \citep{Umetsu2014}, and it is located at a redshift of
$0.439$; at this distance, $1\arcsec$ corresponds to $5.68$ kpc.
It was discovered in the ROSAT All Sky Survey \citep[RXC J1206.2$-$0848;][]{Boehringer2001}.
Even though the cluster appears as a relaxed object \citep{Zitrin2012,Biviano2013}, both the X-ray surface
brightness and the total mass show asymmetric distributions,
which are characterized by a single peak located approximately at the BCG position
(R.A. $=$ 12:06:12.149 and decl. $= -$8:48:03.37).
The multiple-image catalog consists of $82$ images (all spectroscopically
confirmed) from $27$ background sources that span a redshift range from $1.01$
to $6.06$ \citep{Caminha2017b}.
Remarkably, $11$ of these images are in the central $50$ kpc, allowing for a very
accurate measurement of the mass distribution in the core of the cluster.
The total mass model is similar to that by \citet{Caminha2017b} and consists of
three large-scale elliptical halos (DM and intracluster light), three elliptical dPIE
components (hot gas) and $265$ galaxy-scale dPIE halos (member galaxies) and an
external shear.
The large-scale elliptical halos are needed in order to mimic the asymmetric
total mass distribution of the cluster and should not be considered as separate
subclusters.
We choose the luminosity value of the BCG as reference in the scaling relations
of the cluster members.

\begin{table}
\caption{\label{tab:cluster_summary}Summary of cluster properties and lensing
data. For each cluster we report the redshift, $z$, total mass, $M_{200\textrm{c}}$,
radius, $R_{200\textrm{c}}$, number of member galaxy, $N_\textrm{mem}$, and number of secure
multiple images, $N_\textrm{im}$, all spectroscopically confirmed.}
\begin{tabular}{l*{5}{c}}

\hline
\hline
Cluster    & $z$        & $M_{200\textrm{c}}$  & $R_{200\textrm{c}}$  & $N_{\textrm{mem}}$ & $N_{\textrm{im}}$ \\
    &        &  ($10^{15} M_\odot$) & (Mpc) &  & \\
\hline

RXC~J2248  & $0.348$    & $2.03$$\pm$$0.67$     & $2.32$$\pm$$0.26$     & $222$               & $55$ \\
MACS~J0416 & $0.396$    & $1.04$$\pm$$0.22$     & $1.82$$\pm$$0.13$     & $193$               & $102$ \\
MACS~J1206 & $0.439$    & $1.59$$\pm$$0.36$     & $2.06$$\pm$$0.16$     & $265$               & $82$ \\
\hline

\end{tabular}
\end{table}
In Table \ref{tab:cluster_summary}, we summarize for each cluster the most
important information for the lensing and following analysis.

\section{Results}\label{sec:results}
The exquisite quality of the multi-wavelength data, presented in the previous section, allows us to
create very accurate models of the total mass distribution of the galaxy
clusters in the sample.
Here we present these models.
We will only discuss the full models that include both the DM and hot gas
components.
As noted in \citet{Bonamigo2017}, traditional methods can not determine the
model parameters of the DM-only components these are not separated from the hot
gas term, which can only be subtracted a-posteriori.
We refer to \citet{Bonamigo2017} for a more detailed comparison with traditional
techniques.

We measure the cluster temperature and
metallicity by taking the median values of their radial profiles. These have
been derived from the X-ray spectra up to a distance from the cluster center of
$1\arcmin.5$, $4\arcmin.0$, and $3\arcmin.0$, for RXC~J2248, MACS~J0416 and
MACS~J1206, respectively.
The adopted values of temperature (metallicity) are $12.8$ ($0.31$), $10.4$
($0.25$), and $13.0$~KeV ($0.22$).
To describe the hot gas mass of the clusters, we fit the X-ray surface brightness
with three elliptical dPIE profiles (four in the case of MACS~J0416) plus
uniform backgrounds of $0.11$, $0.89$ and $0.02$~counts/pixel, measured
from the \emph{Chandra} images by masking a circular region of radius of approximately
$1.5$~Mpc around each cluster center.
The resulting minimum values of the Cash statistic $C$, used for the fit, are
13794.8 (11551 d.o.f), 3438.5 (2828 d.o.f), and 9785.9 (11589 d.o.f), for
RXC~J2248, MACS~J0416 and MACS~J1206, respectively.
Additionally, we have tried alternative models, which we discuss in Appendix \ref{sec:xray_models}.
\begin{figure*}
	\centering
	\includegraphics[width=\textwidth]{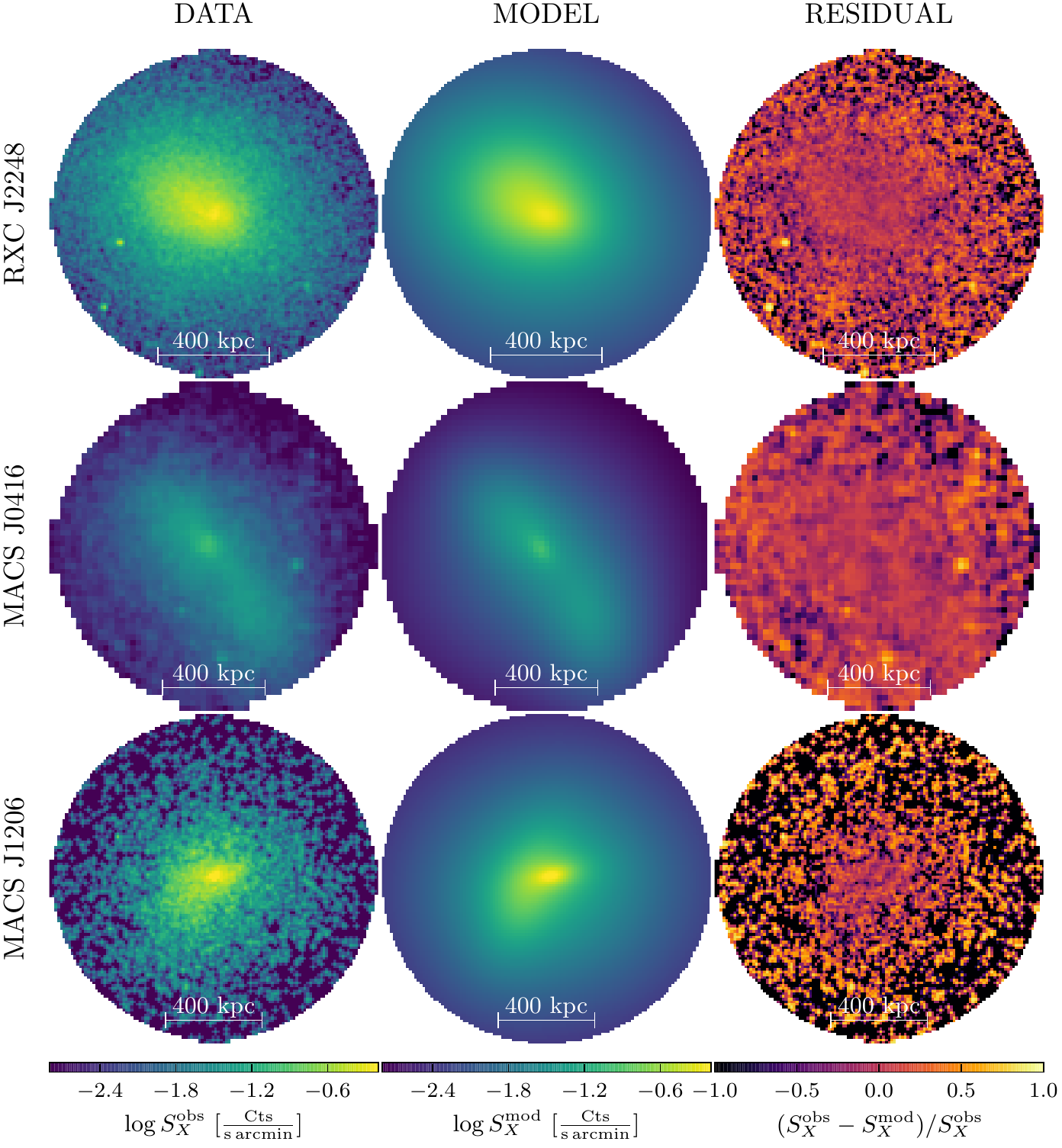}
    \caption{X-ray surface brightness (logarithmic scale) and relative residual
    maps: data (first column), best-fitting model (second column)
	and relative residuals (third column).
	The rows represent, from top to bottom, the clusters RXC~J2248, MACS~J0416 and
	MACS~J1206.
	Each panel shows the circular aperture of radius of $120\arcsec$ used in
	the fitting procedure.
	Point sources are shown only for graphical reasons and have been masked
	out in the fitting procedure.}
	\label{fig:surface_brightness-fit}
\end{figure*}
Figure \ref{fig:surface_brightness-fit} shows the X-ray surface brightness of
the three clusters, one for each row.
The first two columns show the logarithm of the observed data,
$S^\mathrm{obs}_X$, and best-fit model, $S^\mathrm{mod}_X$, respectively.
The images in the third column are the relative difference between the data and
the model.
Only circular apertures of radius $120\arcsec$ are shown, which correspond to
those considered when fitting the cluster X-ray surface brightness.
From these plots, it is clear that the complex geometry of the X-ray surface
brightness requires multiple components to describe either individual
subclusters, like in MACS~J0416, or the asymmetry of the emission, like in
MACS~J1206.
We note that in this work we provide an updated model of the hot intracluster
gas in MACS~J0416 presented in \citet{Bonamigo2017}.
While the previous spherical-component approximation provides a good
fit to the data, the new elliptical model is favored by several model-selection
criteria, such as the Akaike information criterion \citep[AIC,][]{Akaike1974} and
Bayesian Information Criterion \citep[BIC,][]{Schwarz1978}, and it is a more
accurate representation the complex nature of this cluster.
\begin{table}
\centering

\caption{\label{tab:sbfit_best_RXCJ2248}\emph{RXC~J2248}: 
Median values and confidence level (68\%) uncertainties of the Three-Component dPIE Model of the
X-Ray Surface Brightness. The center position refers to the BCG (R.A. $=$ 22:48:43.970
and decl. $= -$44:31:51.16).}
\begin{tabular}{lrrr}

\hline
\hline
 & Comp. 1 & Comp. 2 & Comp. 3\\
\hline

$x_0$ ($\arcsec$) & $27^{ +22 } _{ -21 }$ & $-18.2^{ +0.6 } _{ -0.7 }$ & $0.2^{ +0.4 } _{ -0.4 }$ \\
$y_0$ ($\arcsec$) & $-70^{ +8 } _{ -15 }$ & $13.6^{ +0.5 } _{ -0.5 }$ & $-1.2^{ +0.3 } _{ -0.2 }$ \\
$\epsilon$ & $0.71^{ +0.12 } _{ -0.17 }$ & $0.13^{ +0.01 } _{ -0.01 }$ & $0.34^{ +0.02 } _{ -0.03 }$ \\
$\theta$ (degree) & $-163^{ +8 } _{ -7 }$ & $-29^{ +3 } _{ -2 }$ & $-15^{ +2 } _{ -2 }$ \\
$R_{C}$ ($\arcsec$) & $188.6^{ +0.1 } _{ -0.1 }$ & $36.0^{ +0.5 } _{ -0.6 }$ & $14.6^{ +0.3 } _{ -0.5 }$ \\
$R_{T}$ ($\arcsec$) & $189.1^{ +0.1 } _{ -0.1 }$ & $360^{ +50 } _{ -50 }$ & $359^{ +4 } _{ -5 }$ \\
$\sigma_0$ (km s$^{-1}$) & $440^{ +20 } _{ -30 }$ & $539^{ +6 } _{ -5 }$ & $308^{ +7 } _{ -7 }$ \\
\hline

\end{tabular}

\vspace{20pt}

\caption{\label{tab:sbfit_best_MACSJ0416}\emph{MACS~J0416}: 
Median values and confidence level (68\%) uncertainties of the Four-Component dPIE Model of the
X-Ray Surface Brightness. The center position refers to the northeast BCG, G1
(R.A. $=$ 04:16:09.154 and decl. $= -$24:04:02.90).}
\begin{tabular}{lrrrr}

\hline
\hline
 & Comp. 1 & Comp. 2 & Comp. 3 & Comp. 4 \\
\hline

$x_0$ ($\arcsec$) & $-19.6^{ +2.2 } _{ -1.4 }$ & $30.7^{ +0.5 } _{ -0.7 }$ & $-2.3^{ +0.1 } _{ -0.1 }$ & $-20.2^{ +0.2 } _{ -0.1 }$ \\
$y_0$ ($\arcsec$) & $-13.2^{ +1.4 } _{ -1.1 }$ & $-48.7^{ +0.3 } _{ -0.2 }$ & $-1^{ +0.1 } _{ -0.1 }$ & $14.6^{ +0.6 } _{ -0.4 }$ \\
$\epsilon$ & $0.09^{ +0.05 } _{ -0.04 }$ & $0.41^{ +0.03 } _{ -0.03 }$ & $0.42^{ +0.04 } _{ -0.12 }$ & $0.39^{ +0.03 } _{ -0.02 }$ \\
$\theta$ (degree) & $-160^{ +10 } _{ -20 }$ & $-71^{ +2 } _{ -3 }$ & $-52^{ +5 } _{ -6 }$ & $-47^{ +3 } _{ -2 }$ \\
$R_{C}$ ($\arcsec$) & $149.3^{ +0.1 } _{ -0.1 }$ & $34.8^{ +0.5 } _{ -0.8 }$ & $10.3^{ +0.7 } _{ -0.8 }$ & $50.8^{ +0.9 } _{ -0.9 }$ \\
$R_{T}$ ($\arcsec$) & $149.8^{ +0.1 } _{ -0.1 }$ & $160^{ +20 } _{ -10 }$ & $24.20^{ +7 } _{ -5 }$ & $52.3^{ +0.1 } _{ -0.1 }$ \\
$\sigma_0$ (km s$^{-1}$) & $529^{ +4 } _{ -2 }$ & $306^{ +3 } _{ -4 }$ & $134^{ +6 } _{ -5 }$ & $346^{ +4 } _{ -6 }$ \\
\hline

\end{tabular}

\vspace{20pt}

\caption{\label{tab:sbfit_best_MACSJ1206}\emph{MACS~J1206}: 
Median values and confidence level (68\%) uncertainties of the Three-Component dPIE Model of the
X-Ray Surface Brightness. The center position refers to the BCG (R.A. $=$ 12:06:12.149
and decl. $= -$8:48:03.37).}
\begin{tabular}{lrrrr}

\hline
\hline
 & Comp. 1 & Comp. 2 & Comp. 3 \\
\hline

$x_0$ ($\arcsec$) & $1.7^{ +1.5 } _{ -0.9 }$ & $-13.9^{ +1.5 } _{ -1.8 }$ & $3.3^{ +0.7 } _{ -0.5 }$ \\
$y_0$ ($\arcsec$) & $-6.7^{ +1.5 } _{ -1.3 }$ & $-7.6^{ +1.4 } _{ -1.1 }$ & $2.1^{ +0.4 } _{ -0.1 }$ \\
$\epsilon$ & $0.13^{ +0.03 } _{ -0.06 }$ & $0.50^{ +0.07 } _{ -0.03 }$ & $0.55^{ +0.06 } _{ -0.04 }$ \\
$\theta$ (degree) & $0^{ +8 } _{ -10 }$ & $-111^{ +4 } _{ -5 }$ & $-169^{ +5 } _{ -2 }$ \\
$R_{C}$ ($\arcsec$) & $59^{ +5 } _{ -2 }$ & $39^{ +3 } _{ -2 }$ & $8.3^{ +0.4 } _{ -0.8 }$ \\
$R_{T}$ ($\arcsec$) & $810^{ +180 } _{ -400 }$ & $43.9^{ +0.1 } _{ -0.1 }$ & $200^{ +20 } _{ -110 }$ \\
$\sigma_0$ (km s$^{-1}$) & $536^{ +15 } _{ -6 }$ & $397^{ +22 } _{ -8 }$ & $239^{ +5 } _{ -13 }$ \\
\hline

\end{tabular}

\end{table}
The median values and confidence level (68\%) uncertainties of the mass density
parameters of the intracluster gas are presented in Tables
\ref{tab:sbfit_best_RXCJ2248}, \ref{tab:sbfit_best_MACSJ0416}
and \ref{tab:sbfit_best_MACSJ1206}.
The positions of the centers refer to the BCGs of the clusters (G1 for MACS~J0416).

\begin{table}
\centering
\caption{\label{tab:lensing_RXCJ2248}\emph{RXC~J2248}:
    Median values and confidence level (CL) uncertainties for the strong-lensing
    model parameters. The center position refers to the BCG (R.A. $=$ 22:48:43.970
    and decl. $= -$44:31:51.16). The angle $\theta_{h1}$ is measured counterclockwise from the West axis.}
\begin{tabular}{l*{4}{c}}

\hline
\hline
 & Median & 68\% CL & 95\% CL & 99.7\% CL \\
\hline

$x_{h1}$ ($\arcsec$) & $1.8$ & $^{ +0.3 } _{ -0.3 }$ & $^{ +0.6 } _{ -0.6 }$ & $^{ +1.0 } _{ -0.8 }$ \\
$y_{h1}$ ($\arcsec$) & $-1.0$ & $^{ +0.2 } _{ -0.2 }$ & $^{ +0.4 } _{ -0.5 }$ & $^{ +0.6 } _{ -0.7 }$ \\
$\epsilon_{h1}$ & $0.64$ & $^{ +0.01 } _{ -0.01 }$ & $^{ +0.02 } _{ -0.01 }$ & $^{ +0.03 } _{ -0.02 }$ \\
$\theta_{h1}$ (degree) & $-38.7$ & $^{ +0.3 } _{ -0.3 }$ & $^{ +0.5 } _{ -0.5 }$ & $^{ +0.7 } _{ -0.8 }$ \\
$R_{C,h1}$ ($\arcsec$) & $19.2$ & $^{ +0.7 } _{ -0.6 }$ & $^{ +1.3 } _{ -1.3 }$ & $^{ +2.0 } _{ -1.9 }$ \\
$\sigma_{0,h1}$ (km s$^{-1} $) & $1396$ & $^{ +15 } _{ -17 }$ & $^{ +27 } _{ -35 }$ & $^{ +38 } _{ -53 }$ \\
\hline
$x_{h2}$ ($\arcsec$) & $-53.3$ & $^{ +4.8 } _{ -6.0 }$ & $^{ +8.8 } _{ -12.8 }$ & $^{ +12.0 } _{ -21.4 }$ \\
$y_{h2}$ ($\arcsec$) & $27.1$ & $^{ +2.8 } _{ -2.5 }$ & $^{ +6.3 } _{ -4.8 }$ & $^{ +10.9 } _{ -7.1 }$ \\
$\sigma_{0,h2}$ (km s$^{-1} $) & $282$ & $^{ +35 } _{ -31 }$ & $^{ +74 } _{ -60 }$ & $^{ +118 } _{ -86 }$ \\
\hline
$R_{T,g}$ ($\arcsec$) & $46.6$ & $^{ +17.8 } _{ -13.8 }$ & $^{ +42.0 } _{ -25.1 }$ & $^{ +72.8 } _{ -33.7 }$ \\
$\sigma_{0,g}$ (km s$^{-1} $) & $274$ & $^{ +16 } _{ -16 }$ & $^{ +32 } _{ -33 }$ & $^{ +53 } _{ -51 }$ \\
\hline

\end{tabular}
\end{table}
\begin{table}
\centering
\caption{\label{tab:lensing_MACSJ0416}\emph{MACS~J0416}:
    Median values and confidence level (CL) uncertainties for the strong-lensing
    model parameters. The center position refers to the northeast BCG, G1
    (R.A. $=$ 04:16:09.154 and decl. $= -$24:04:02.90). The angles $\theta_{h1}$
    and $\theta_{h2}$ are measured counterclockwise from the West axis.}
\begin{tabular}{l*{4}{c}}

\hline
\hline
 & Median & 68\% CL & 95\% CL & 99.7\% CL \\
\hline

$x_{h1}$ ($\arcsec$) & $-2.5$ & $^{ +0.9 } _{ -0.8 }$ & $^{ +1.8 } _{ -1.4 }$ & $^{ +2.8 } _{ -1.9 }$ \\
$y_{h1}$ ($\arcsec$) & $1.8$ & $^{ +0.5 } _{ -0.6 }$ & $^{ +0.9 } _{ -1.2 }$ & $^{ +1.3 } _{ -2.0 }$ \\
$\epsilon_{h1}$ & $0.86$ & $^{ +0.01 } _{ -0.01 }$ & $^{ +0.03 } _{ -0.03 }$ & $^{ +0.04 } _{ -0.06 }$ \\
$\theta_{h1}$ (degree) & $145.2$ & $^{ +0.7 } _{ -0.9 }$ & $^{ +1.5 } _{ -1.9 }$ & $^{ +2.6 } _{ -2.8 }$ \\
$R_{C,h1}$ ($\arcsec$) & $6.7$ & $^{ +0.7 } _{ -0.9 }$ & $^{ +1.4 } _{ -1.8 }$ & $^{ +2.1 } _{ -2.5 }$ \\
$\sigma_{0,h1}$ (km s$^{-1} $) & $707$ & $^{ +23 } _{ -23 }$ & $^{ +45 } _{ -51 }$ & $^{ +64 } _{ -83 }$ \\
\hline
$x_{h2}$ ($\arcsec$) & $19.9$ & $^{ +0.3 } _{ -0.3 }$ & $^{ +0.8 } _{ -0.7 }$ & $^{ +1.6 } _{ -1.1 }$ \\
$y_{h2}$ ($\arcsec$) & $-37.0$ & $^{ +0.6 } _{ -0.6 }$ & $^{ +1.3 } _{ -1.4 }$ & $^{ +1.9 } _{ -2.6 }$ \\
$\epsilon_{h2}$ & $0.77$ & $^{ +0.01 } _{ -0.01 }$ & $^{ +0.03 } _{ -0.03 }$ & $^{ +0.04 } _{ -0.05 }$ \\
$\theta_{h2}$ (degree) & $126.1$ & $^{ +0.4 } _{ -0.4 }$ & $^{ +0.8 } _{ -0.9 }$ & $^{ +1.3 } _{ -1.3 }$ \\
$R_{C,h2}$ ($\arcsec$) & $12.5$ & $^{ +0.6 } _{ -0.7 }$ & $^{ +1.2 } _{ -1.5 }$ & $^{ +1.9 } _{ -2.2 }$ \\
$\sigma_{0,h2}$ (km s$^{-1} $) & $1064$ & $^{ +16 } _{ -17 }$ & $^{ +31 } _{ -37 }$ & $^{ +49 } _{ -63 }$ \\
\hline
$x_{h3}$ ($\arcsec$) & $-34.4$ & $^{ +0.9 } _{ -1.1 }$ & $^{ +1.9 } _{ -2.6 }$ & $^{ +2.7 } _{ -4.6 }$ \\
$y_{h3}$ ($\arcsec$) & $8.1$ & $^{ +1.0 } _{ -0.7 }$ & $^{ +2.7 } _{ -1.4 }$ & $^{ +4.3 } _{ -2.0 }$ \\
$R_{C,h3}$ ($\arcsec$) & $4.4$ & $^{ +2.4 } _{ -2.2 }$ & $^{ +4.8 } _{ -3.8 }$ & $^{ +7.4 } _{ -4.3 }$ \\
$\sigma_{0,h3}$ (km s$^{-1} $) & $350$ & $^{ +51 } _{ -48 }$ & $^{ +104 } _{ -84 }$ & $^{ +167 } _{ -107 }$ \\
\hline
$R_{T,g}$ ($\arcsec$) & $7.8$ & $^{ +2.3 } _{ -1.5 }$ & $^{ +7.7 } _{ -3.4 }$ & $^{ +12.6 } _{ -4.5 }$ \\
$\sigma_{0,g}$ (km s$^{-1} $) & $318$ & $^{ +11 } _{ -74 }$ & $^{ +42 } _{ -101 }$ & $^{ +67 } _{ -121 }$ \\
\hline

\end{tabular}
\end{table}
\begin{table}
\centering
\caption{\label{tab:lensing_MACSJ1206}\emph{MACS~J1206}:
    Median values and confidence level (CL) uncertainties for the strong-lensing
    model parameters. The center position refers to the BCG (R.A. $=$ 12:06:12.149
    and decl. $= -$8:48:03.37). The angles $\theta_{h1}$, $\theta_{h2}$,
    $\theta_{h3}$, and $\theta_{4}$ are measured counterclockwise from the West
    axis. The parameters $\gamma_4$ and $\theta_4$ are the shear and its angle.}
\begin{tabular}{l*{4}{c}}

\hline
\hline
 & Median & 68\% CL & 95\% CL & 99.7\% CL \\
\hline

$x_{h1}$ ($\arcsec$) & $-0.9$ & $^{ +0.4 } _{ -0.5 }$ & $^{ +0.8 } _{ -1.0 }$ & $^{ +1.2 } _{ -1.4 }$ \\
$y_{h1}$ ($\arcsec$) & $0.3$ & $^{ +0.2 } _{ -0.3 }$ & $^{ +0.4 } _{ -0.5 }$ & $^{ +0.6 } _{ -0.8 }$ \\
$\epsilon_{h1}$ & $0.69$ & $^{ +0.03 } _{ -0.03 }$ & $^{ +0.05 } _{ -0.05 }$ & $^{ +0.08 } _{ -0.09 }$ \\
$\theta_{h1}$ (degree) & $19.7$ & $^{ +0.9 } _{ -0.8 }$ & $^{ +1.8 } _{ -1.7 }$ & $^{ +2.7 } _{ -2.6 }$ \\
$R_{C,h1}$ ($\arcsec$) & $6.7$ & $^{ +0.5 } _{ -0.5 }$ & $^{ +1.2 } _{ -1.1 }$ & $^{ +2.4 } _{ -1.6 }$ \\
$\sigma_{0,h1}$ (km s$^{-1} $) & $968$ & $^{ +41 } _{ -46 }$ & $^{ +95 } _{ -98 }$ & $^{ +182 } _{ -146 }$ \\
\hline
$x_{h2}$ ($\arcsec$) & $9.5$ & $^{ +0.8 } _{ -0.7 }$ & $^{ +1.8 } _{ -1.4 }$ & $^{ +2.9 } _{ -2.1 }$ \\
$y_{h2}$ ($\arcsec$) & $4.0$ & $^{ +0.7 } _{ -0.7 }$ & $^{ +1.5 } _{ -1.8 }$ & $^{ +2.4 } _{ -3.8 }$ \\
$\epsilon_{h2}$ & $0.55$ & $^{ +0.10 } _{ -0.11 }$ & $^{ +0.20 } _{ -0.21 }$ & $^{ +0.32 } _{ -0.30 }$ \\
$\theta_{h2}$ (degree) & $115.1$ & $^{ +3.0 } _{ -2.4 }$ & $^{ +6.5 } _{ -4.5 }$ & $^{ +10.3 } _{ -6.4 }$ \\
$R_{C,h2}$ ($\arcsec$) & $13.9$ & $^{ +1.5 } _{ -1.1 }$ & $^{ +3.4 } _{ -2.0 }$ & $^{ +6.1 } _{ -2.8 }$ \\
$\sigma_{0,h2}$ (km s$^{-1} $) & $758$ & $^{ +37 } _{ -36 }$ & $^{ +79 } _{ -79 }$ & $^{ +123 } _{ -147 }$ \\
\hline
$x_{h3}$ ($\arcsec$) & $-28.6$ & $^{ +1.4 } _{ -1.7 }$ & $^{ +2.8 } _{ -5.1 }$ & $^{ +4.2 } _{ -26.2 }$ \\
$y_{h3}$ ($\arcsec$) & $-6.7$ & $^{ +0.9 } _{ -0.8 }$ & $^{ +2.8 } _{ -1.6 }$ & $^{ +7.6 } _{ -2.5 }$ \\
$\epsilon_{h3}$ & $0.35$ & $^{ +0.06 } _{ -0.06 }$ & $^{ +0.13 } _{ -0.13 }$ & $^{ +0.19 } _{ -0.20 }$ \\
$\theta_{h3}$ (degree) & $-25.4$ & $^{ +10.2 } _{ -11.6 }$ & $^{ +18.0 } _{ -22.5 }$ & $^{ +24.0 } _{ -33.7 }$ \\
$R_{C,h3}$ ($\arcsec$) & $12.3$ & $^{ +2.3 } _{ -2.1 }$ & $^{ +5.1 } _{ -4.0 }$ & $^{ +14.1 } _{ -6.1 }$ \\
$\sigma_{0,h3}$ (km s$^{-1} $) & $600$ & $^{ +45 } _{ -41 }$ & $^{ +97 } _{ -84 }$ & $^{ +157 } _{ -159 }$ \\
\hline
$\gamma_{4}$ & $0.11$ & $^{ +0.01 } _{ -0.01 }$ & $^{ +0.02 } _{ -0.02 }$ & $^{ +0.03 } _{ -0.03 }$ \\
$\theta_{4}$ (degree) & $101.5$ & $^{ +1.5 } _{ -1.4 }$ & $^{ +3.0 } _{ -3.4 }$ & $^{ +4.8 } _{ -6.8 }$ \\
\hline
$R_{T,g}$ ($\arcsec$) & $3.6$ & $^{ +0.9 } _{ -0.7 }$ & $^{ +1.9 } _{ -1.4 }$ & $^{ +3.2 } _{ -2.0 }$ \\
$\sigma_{0,g}$ (km s$^{-1} $) & $353$ & $^{ +24 } _{ -21 }$ & $^{ +53 } _{ -40 }$ & $^{ +93 } _{ -57 }$ \\
\hline

\end{tabular}
\end{table}
The next step in the analysis is to include these hot gas models as fixed mass
components in the strong lensing analysis.
We use a first optimization of the lensing model of each cluster to derive new values for the
error in the positions of the multiple images.
These are $0\arcsec.46$, $0\arcsec.57$ and $0\arcsec.35$, for RXC~J2248,
MACS~J0416 and MACS~J1206, respectively.
The subsequent best-fit models have values of the minimum-$\chi^2$ of $59.71$ (59 d.o.f),
$111.0$ (110  d.o.f) and $90.9$ (88  d.o.f); the corresponding values of the rms
of the multiple-image position offsets are $0\arcsec.48$, $0\arcsec.59$ and
$0\arcsec.45$ (median values $0\arcsec.37$, $0\arcsec.40$ and $0\arcsec.36$).
Tables \ref{tab:lensing_RXCJ2248}, \ref{tab:lensing_MACSJ0416} and
\ref{tab:lensing_MACSJ1206} contain the inferred values of the mass model
parameters\footnote{The \emph{lenstool} input files and sampled posterior
distributions can be found at \url{https://sites.google.com/site/vltclashpublic/}}.
Here, we quote the median values and the $68\%$, $95\%$ and $99.7\%$
confidence level (CL) intervals.

\begin{figure*}
	\centering
	\includegraphics[width=0.75\textwidth]{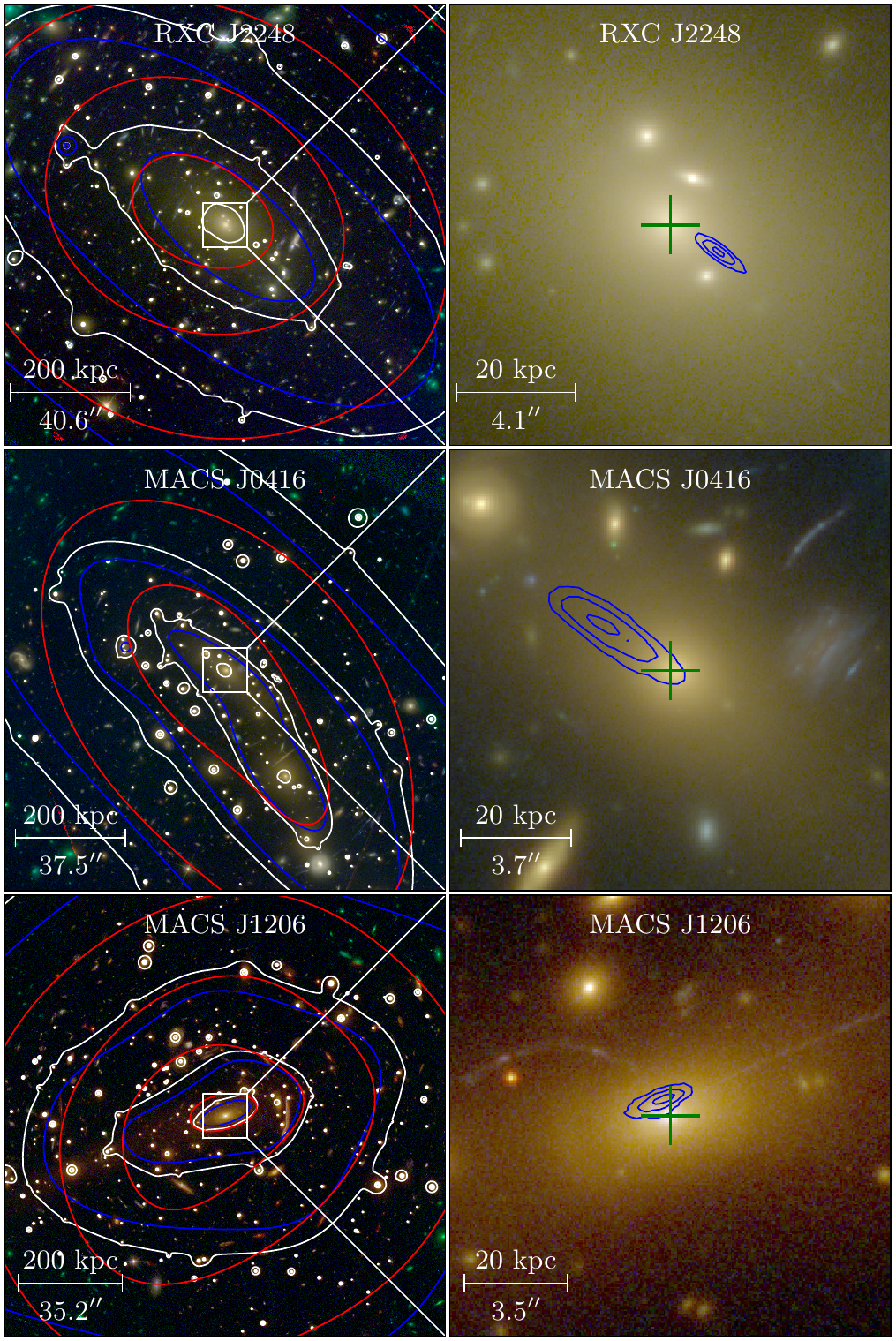}
    \caption{Surface mass density of RXC~J2248 (top row), MACS~J0416 (middle
    row) and MACS~J1206 (bottom row) overlaid on color-composite HST WCF3+ACS
    images. White, blue and red isodensity contours (left panel) correspond to total,
    diffuse DM and hot intracluster gas, respectively. The total and diffuse DM
    contours are drawn at $3.5$, $7.2$, $15$ and $30$ in units of
    $10^{-4}$ M$_\odot/$kpc$^2$; while the hot gas contours are drawn at $0.65$,
    $1.0$, $1.6$ and $2.5$ in units of $10^{-4}$ M$_\odot/$kpc$^2$.
    The right panel is a zoom-in of the central region of each cluster.
    The green plus sign indicates the position of the BCG and has a total width
    of $2\arcsec$. Blue contours (right panel) show the $1\sigma$, $2\sigma$
    and $3\sigma$ confidence levels of the diffuse DM component density peak.}
	\label{fig:components_map}
\end{figure*}
The surface mass densities of the different cluster components are shown in Figure
\ref{fig:components_map} (left panel), where they are represented as
iso-contours overlaid on color-composite HST WCF3/ACS images of the clusters.
The total (white) and DM (blue) isodensity contours are drawn at $3.5$,
$7.2$, $15$ and $30$ in units of $10^{-4}$ M$_\odot/$kpc$^2$; while the gas
(red) isodensity contours are drawn at $0.65$, $1.0$, $1.6$ and $2.5$ in units of
$10^{-4}$ M$_\odot/$kpc$^2$.
The right panel of Figure \ref{fig:components_map} shows a zoom-in of the
central region of the images on the left.
The positions of the BCGs are shown with green plus signs, that are $2\arcsec$
of width.
The blue contours in the right panel show the position of the diffuse DM component
density peak with $1\sigma$, $2\sigma$ and $3\sigma$ confidence levels.
From these plots, the need for a full component separation is clear: the DM and
hot-gas mass distributions have different shapes and centers, due to their
intrinsically different physical properties.
For example, while the DM component in RXC~J2248 is roughly centered on the BCG, the
hot-gas mass distribution is skewed towards northeast.
In MACS~J1206, this latter component is elongated towards the southeast region of the
cluster, creating a twist in the iso-density contours; such feature is not
present in the DM mass distribution, which, however, tends to be fairly lopsided.
In all clusters, especially in RXC~J2248, the hot-gas component is rounder than
the DM one; in MACS~J0416, this happens a-symmetrically, with the northeast
region being rounder than the southeast one.
Moreover, with the exception of RXC~J2248, we find that the peaks of the density of the
diffuse DM components are consistent, within 3$\sigma$, with the positions of the BCGs.
Indeed, the distances between the BCGs and the DM component density peaks are
$(9.3^{ +1.7 } _{ -1.7 })$, $(13.8^{ +4.7 }_{ -5.3 })$, and $(3.9^{ +1.0 }_{ -0.8 })$~kpc,
for RXC~J2248, MACS~J0416 and MACS~J1206, respectively.
We note that RXC~J2248 is described by the simplest DM
mass model and that the DM component seems to counterbalance the a-symmetric hot gas
component, which is skewed in the opposite direction.

\begin{figure*}
	\centering
	\includegraphics[width=\textwidth]{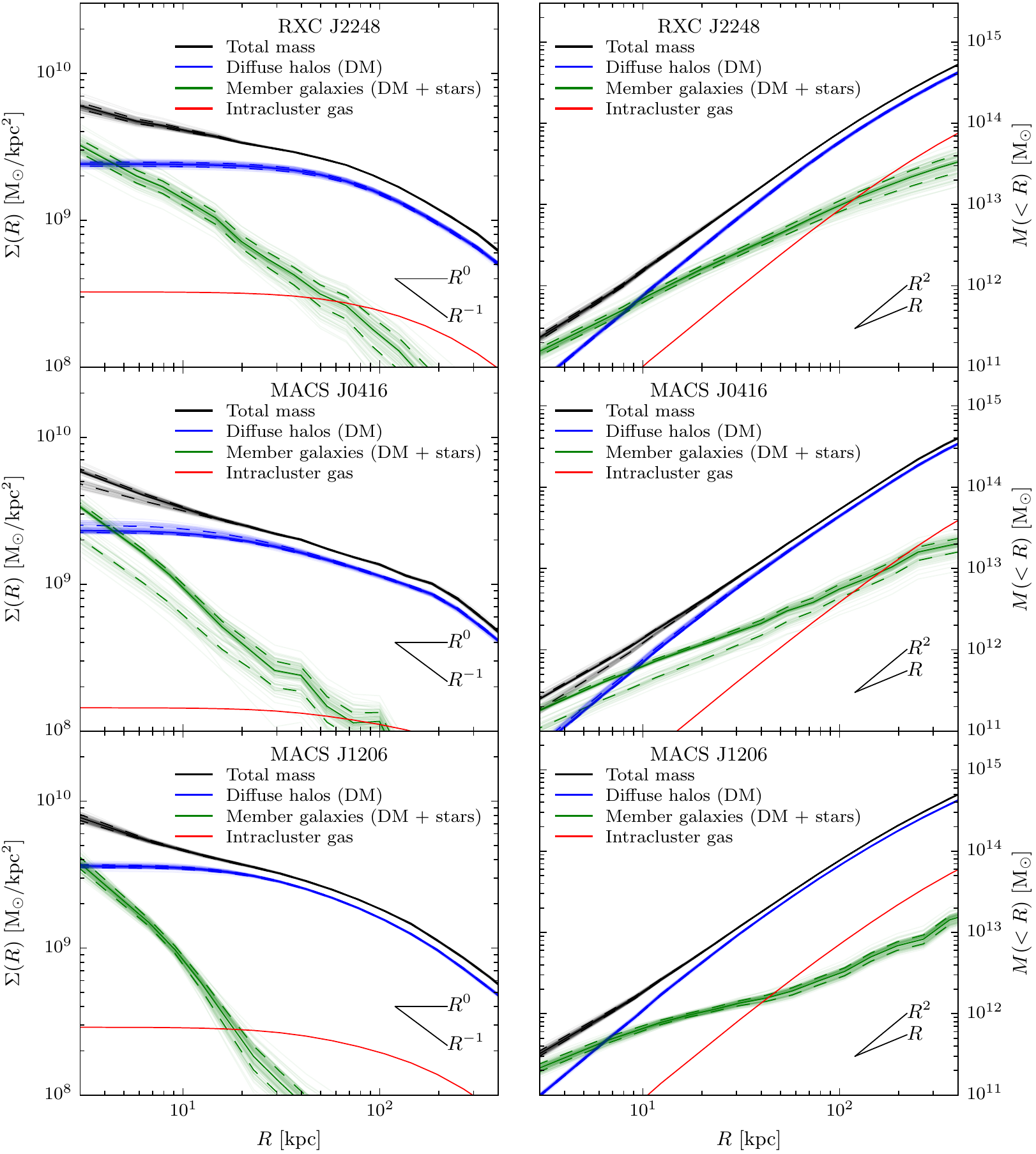}
	    \caption{Radial profiles of the surface mass density (left panel) and
        cumulative projected mass (right panel) for the different components of RXC~J2248
	    (first row), MACS~J0416 (second row) and MACS~J1206 (third row).
	    Black, blue, red, and green curves correspond to total, diffuse DM, hot
	    gas and cluster members, respectively. Thin lines show a subsample of
	    the posterior distribution, while solid and dashed lines show the median
	    and $16^\text{th}\,$-$\,84^\text{th}$ percentiles. Reference logarithmic
	    slopes are also indicated.}
	\label{fig:mass_decomposition}
\end{figure*}
From the previous maps, we obtain the radial profiles shown in Figure
\ref{fig:mass_decomposition}.
Left and right panels contain, respectively, the surface mass density and cumulative projected
mass profiles; each color corresponds to a different mass component: cluster
members (green), hot intracluster gas (red), DM (blue), and total (black).
Thin lines represent a subsample of the models extracted from the sampling of
the posterior distribution, while solid and dashed lines show their median and
$16^\text{th}\,$-$\,84^\text{th}$ percentiles, respectively.

\section{Discussion}\label{sec:discussion}
In order to compare the results of the different clusters in a consistent way,
we decide to rescale the values of their masses, surface mass densities, and
radii.
To do this, we use the values of the mass $M_{200\textrm{c}}$ and of the corresponding
radius $R_{200\textrm{c}}$, derived by \citet{Umetsu2014} via a weak-lensing shear-and-magnification
analysis (see Table \ref{tab:cluster_summary}).
Respectively, for RXC~J2248, MACS~J0416 and MACS~J1206 the values of $M_{200\textrm{c}}$
are $(2.03 \pm 0.67)$, $(1.04 \pm 0.22)$ and $(1.59 \pm 0.36) \times 10^{15}$ M$_\odot$.
In passing, we mention that these values are consistent, given the errors, with
those obtained by \citet{Biviano2013} and \citet{Balestra2016} from the
dynamical analyses of MACS J1206 and MACS J0416, respectively.
These masses correspond to values of $R_{200\textrm{c}}$ of
$(2.32 \pm 0.26)$, $(1.82 \pm 0.13)$ and $(2.06 \pm 0.16)$ Mpc.
In our analysis, we include the error of the weak-lensing measurement of a cluster
total mass, $M_{200\textrm{c}}$, and radius, $R_{200\textrm{c}}$, by considering
that they are described by Gaussian distributions.
\begin{figure*}
	\centering
	\includegraphics[width=\textwidth]{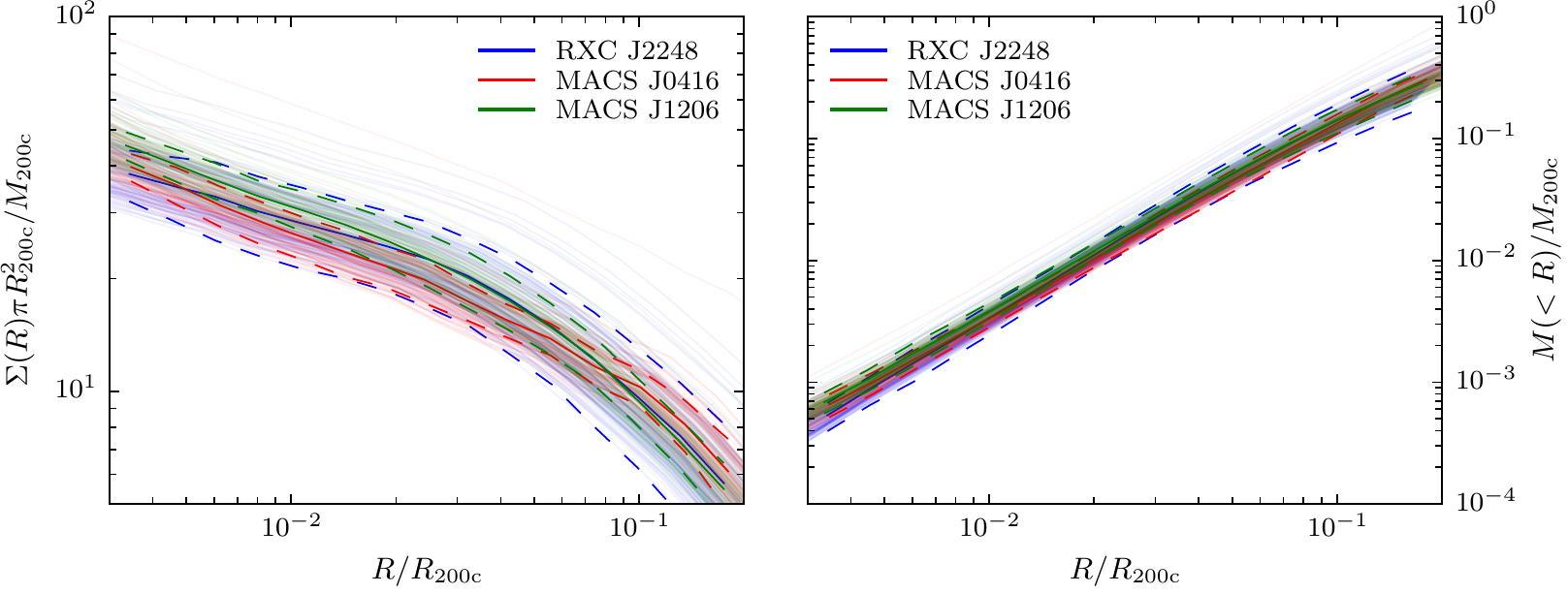}
    \caption{Normalized radial profiles of the total surface mass density (left panel) and
        cumulative projected mass (right panel) of RXC~J2248
	    (blue), MACS~J0416 (red) and MACS~J1206 (green). Thin lines show a
	    subsample of the posterior distribution, while solid and dashed lines
	    show the mean and standard deviation.}
	\label{fig:profile_clusters}
\end{figure*}
Once rescaled, the total surface mass density and cumulative projected mass
profiles show a very uniform behavior, as it can be seen on the left and right
panels of Figure \ref{fig:profile_clusters}, respectively.
Thin lines represent a subsample of the models extracted from the sampling,
while solid and dashed lines show the mean and standard deviation.
In these and following figures, we have identified each cluster with a
different color: blue for RXC~J2248, red for MACS~J0416 and green for MACS~J1206.
Even though these measurements are dominated by the errors on the total mass and
radius from the weak-lensing analysis, the differences between the three clusters
are small, both in terms of density and enclosed mass, suggesting the existence
of a homologous mass profile.
In Appendix \ref{sec:nfw_profiles}, we fit these profiles with
Navarro-Frenk-White profiles \citep{Navarro1997} and discuss the caveats of
claiming ``universality'' of mass profiles that are fitted in projection.
It should be noted that this remarkably good agreement between the rescaled mass
profiles was not expected a priori: the values of $M_{200\textrm{c}}$ have been measured
at much larger scales ($R \sim R_{200\textrm{c}}$) through weak lensing \citep{Umetsu2014}, while the
current analysis is restricted only to the cores of the clusters
($R < 0.2 R_{200\textrm{c}}$).
Previous works \citep{Biviano2013,Grillo2015,Balestra2016,Caminha2017b} have
shown though that in these clusters with high-quality data the results of
different mass diagnostics, where they overlap, agree very well.

Using the surface mass densities of the different cluster components (from Figure
\ref{fig:components_map}), we can create maps of the diffuse DM and hot gas
over total mass fractions.
\begin{figure*}
	\centering
	\includegraphics[width=\columnwidth]{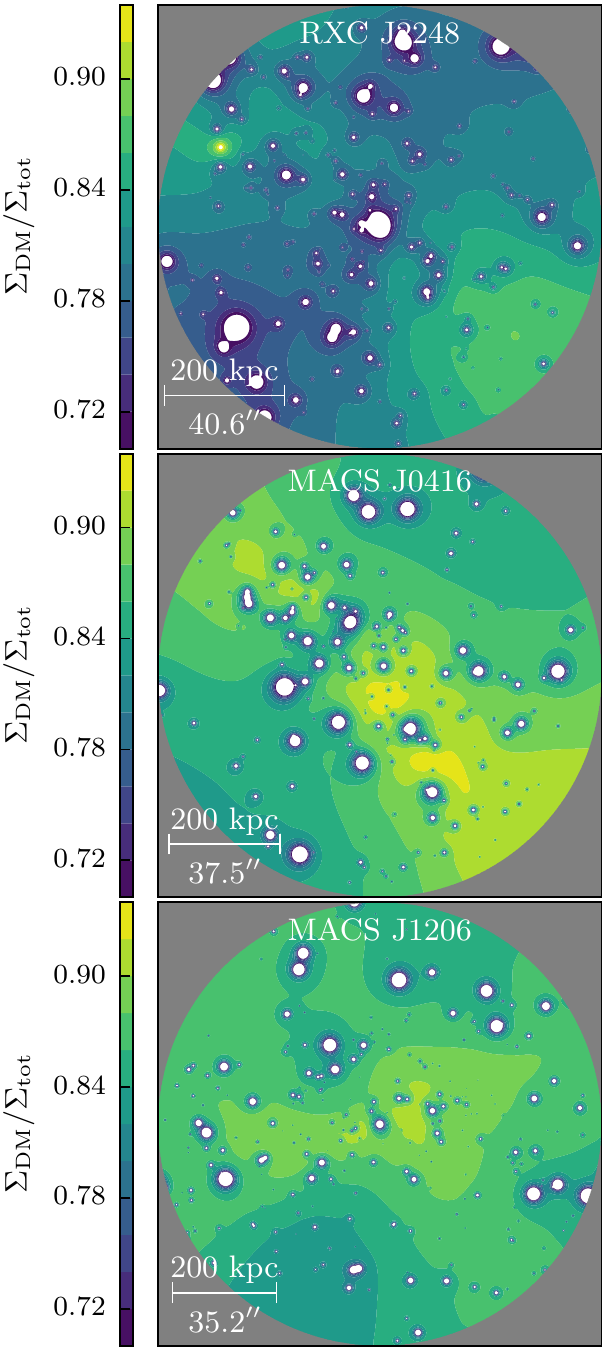}
	\includegraphics[width=\columnwidth]{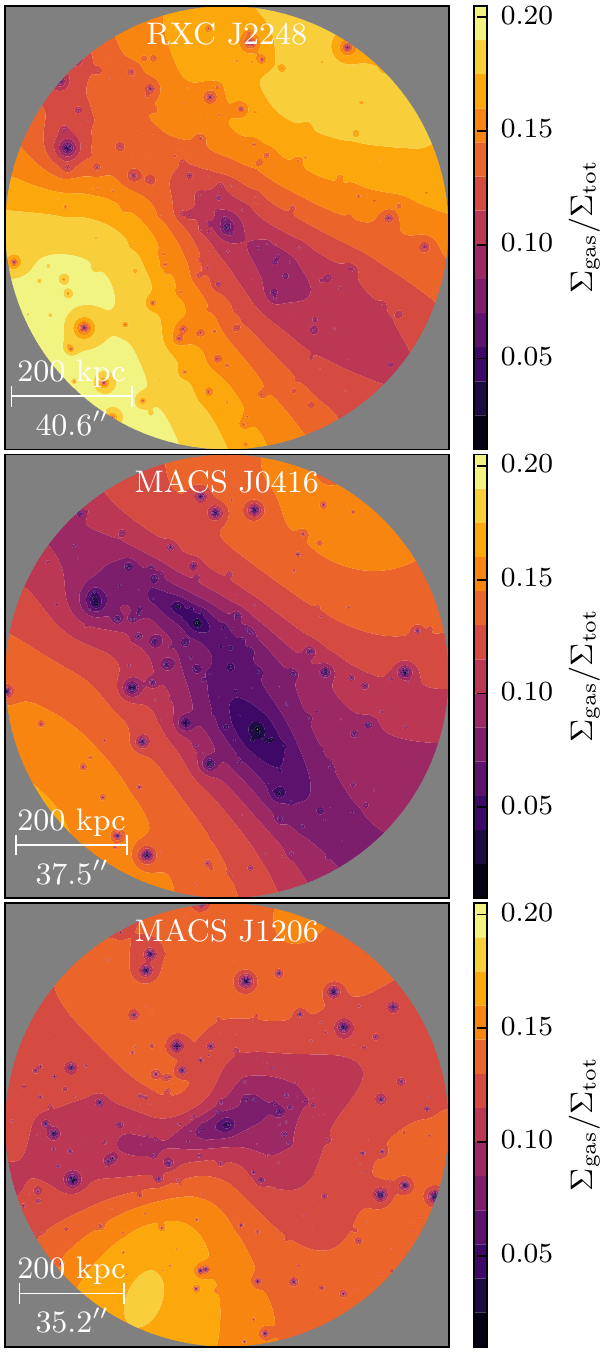}
    \caption{Maps of the diffuse DM (left column) and hot intracluster gas
    (right column) over total mass fractions of RXC~J2248 (first row), MACS~J0416 (second row) and
    MACS~J1206 (third row).}
	\label{fig:ratio_map}
\end{figure*}
These are shown, respectively, on the left and right panel of Figure \ref{fig:ratio_map}.
While the galaxy clusters MACS~J0416 and MACS~J1206 have comparable fractions of DM,
with similar trend with the radial distance from the center,
RXC~J2248 has a lower DM fraction and a much steeper radial dependence.
Moreover, the  alignment of the DM fraction iso-contours seems to be perpendicular to the
orientation of the cluster.
This apparent misalignment is due to a combination of multiple factors.
First, the difference in shape between the DM and hot gas mass
distributions causes the DM to become less dominant along the direction
perpendicular to its mass distribution major axis; although less pronounced, this
effect can also be seen in MACS~J0416.
Secondly, the offset between the gas and DM components causes the southeast
region to have less hot gas and therefore a larger DM mass fraction.
Moreover, these effects are increased by the larger truncation radii
of the cluster member galaxies, $R_{T,g}$, that extend their influence in the
total mass budget.
On the other hand, the hot gas over total mass fraction maps are more consistent
between the clusters, with their centers having lower fractions than the
outskirts.
However, the central regions of MACS~J0416 show a low value, which is consistent
with the cluster merging nature: the turbulence of the merger is heating the
gas, preventing it to fall to the center.
In contrast, RXC~J2248 and MACS~J1206 have a cool-core, that results in the 
higher gas fraction shown on the left panel of Figure \ref{fig:ratio_map}.
\begin{figure}
	\centering
	\includegraphics[width=\columnwidth]{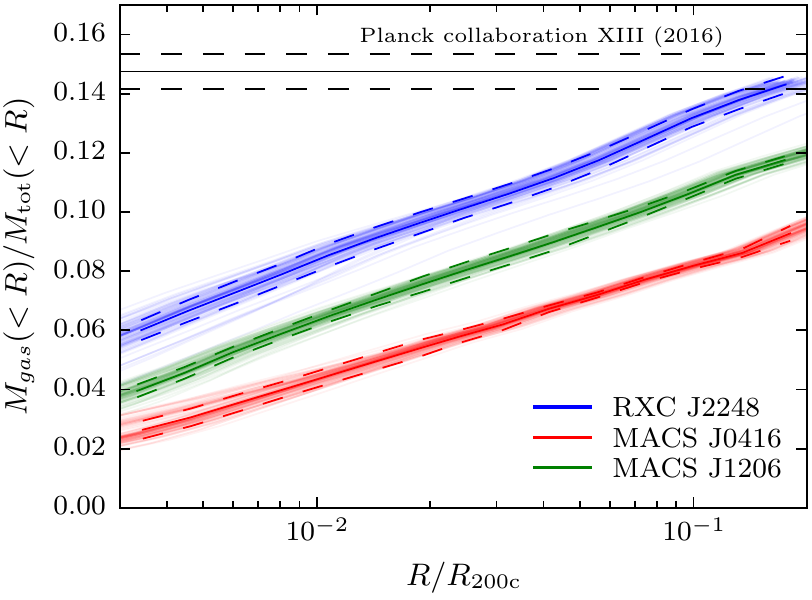}
    \caption{Ratios of the cumulative projected hot gas and total mass for
    RXC~J2248 (blue), MACS~J0416 (red) and MACS~J1206 (green). Thin lines show a
	    subsample of the posterior distribution, while solid and dashed lines
	    show the mean and standard deviation.}
	\label{fig:gas_fraction}
\end{figure}
The same trend can be seen in Figure \ref{fig:gas_fraction}, where we show
the ratios of the cumulative projected hot gas and total mass.
As before, we rescale the radial distance by $R_{200\textrm{c}}$, to
remove the dependence on the cluster mass, and include the corresponding 
uncertainties which dominate the errors shown in Figure \ref{fig:gas_fraction}.
Colors and lines are defined as in Figure \ref{fig:profile_clusters}.
The horizontal black solid and dashed lines show the cosmological baryon
fraction with errors measured by the \emph{Planck} satellite \citepalias{PlanckXIII}.
This is the quantity of cosmological interest, which can be obtained by
adding the fraction of stellar mass to that of the hot gas
measured in this paper, as done by \citet{Annunziatella2017} in MACS~J0416.
We postpone to future works the measurement of the stellar mass component of RXC~J2248 and
MACS~J1206.
Noticeably, we are able to measure the hot gas fraction with very high precision,
(less than $1\%$ before rescaling), thanks to the small uncertainties on the cluster total mass
derived from our high-precision strong lensing models.

Finally, expanding the analysis by \citet{Grillo2015} and \citet{Munari2016}, we look at the substructure
statistics in the three clusters, traced by their member galaxies.
We compare our measurements, derived from the new strong lensing model
(Section \ref{sec:lensing}), to those presented by \citet{Grillo2015}, both on
the observed data of MACS~J0416 and the simulated halos.
In order to increase the statistics, we consider here simulated ($N$-body)
halos that have $M_{200c}$ values larger than $9 \times 10^{14} M_\odot$ from
four snapshots, at the following redshifts: $0.25$, $0.28$, $0.46$ and $0.51$. 
In all clusters, we select only those galaxies, or sub-halos, that have a circular
velocity value $v_c = \sqrt{2} \sigma_0 $ larger than $90$ km s$^{-1}$ and that
are located within a projected distance of $0.16\times R_{200\mathrm{c}}$ from
the cluster centers.
Compared to the previous studies, we change these velocity and projected
distance limits in order to better compare clusters with
different masses and to avoid the mass resolution limit that would otherwise
contaminate the results at the low circular velocity end for the simulated sub-halos.
We refer to the original papers for any other details on the strong lensing model
and simulated data.
\begin{figure*}
	\centering
	\includegraphics[width=\textwidth]{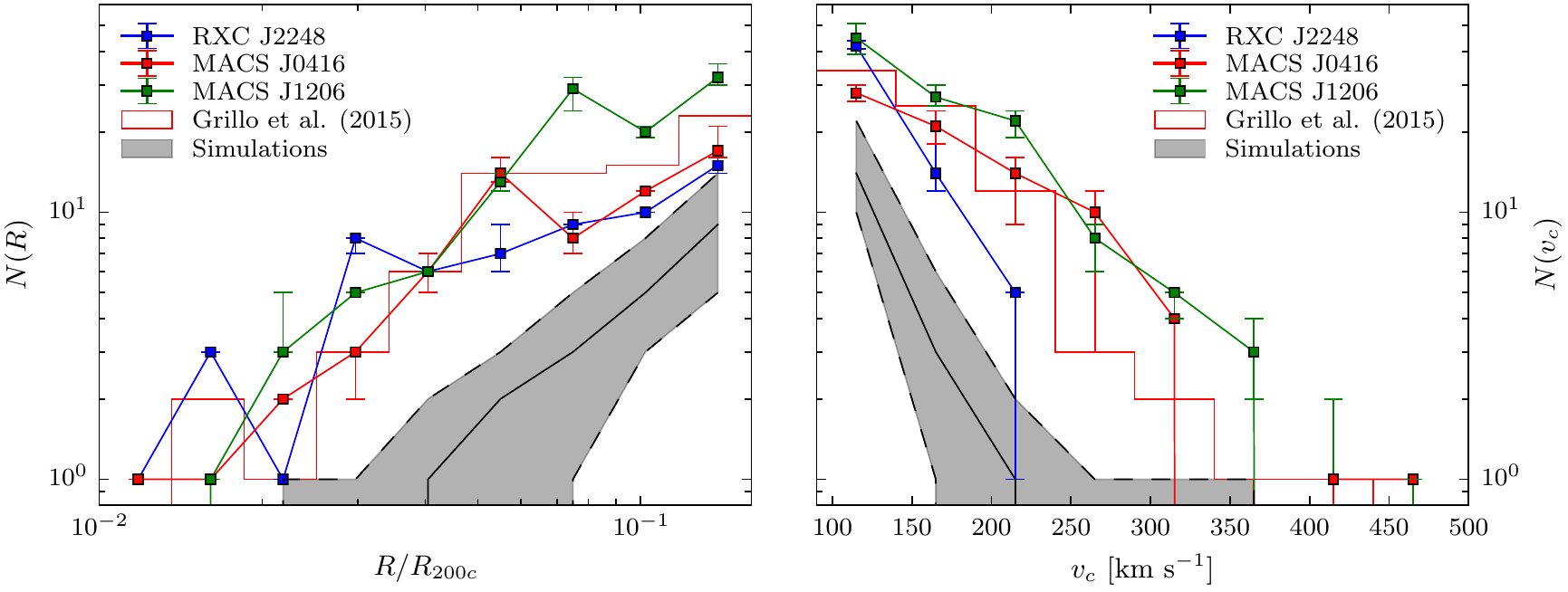}
    \caption{Radial profiles of the number density and circular velocity
    distributions of the member galaxies for RXC~J2248 (blue), MACS~J0416 (red) and
    MACS~J1206 (green). Only galaxies within a circular aperture of $0.16\times R_{200\mathrm{c}}$ 
    and with circular velocity larger than $90$~km~s$^{-1}$ are considered. Points and error bars show the median and
    $16^\text{th}\,$-$\,84^\text{th}$ percentiles, respectively.
    Red histograms are the values derived from the model of MACS~J0416 by
    \citet{Grillo2015}, where a slightly different value for the center of the
    cluster was adopted.}
	\label{fig:subhalo_distributions}
\end{figure*}
Then, we compute the galaxy radial number distribution and circular velocity
distribution.
We show them, respectively, in the left and right panels of Figure
\ref{fig:subhalo_distributions}.
\begin{figure}
	\centering
	\includegraphics[width=\columnwidth]{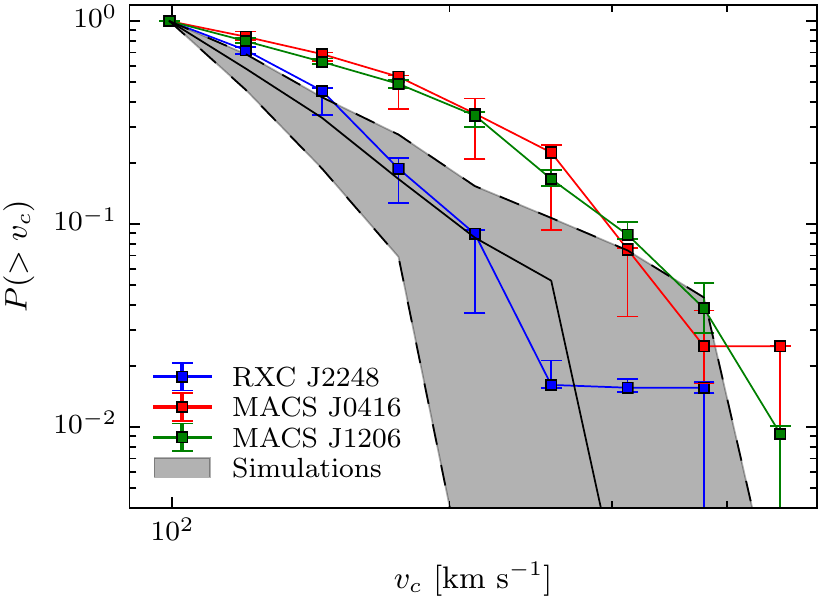}
    \caption{Circular velocity function of the member galaxies for RXC~J2248 (blue), MACS~J0416 (red) and
    MACS~J1206 (green). Only galaxies within a circular aperture of $0.16\times R_{200\mathrm{c}}$ 
    and with a circular velocity larger than $90$~km~s$^{-1}$ are considered. Points and error bars show the median and
    $16^\text{th}\,$-$\,84^\text{th}$ percentiles, respectively.}
	\label{fig:circular_velocity_function}
\end{figure}
Moreover, in Figure \ref{fig:circular_velocity_function}, we show the circular
velocity function of the member galaxies, i.e.\ the fraction of member galaxies
with circular velocity larger than the considered value.
In both plots, the points with error bars mark the median and $16^\text{th}\,$-$\,84^\text{th}$
percentiles, computed from a sample of models extracted from the MCMC sampling.
The gray-shaded areas represent the values obtained from numerical
simulations, as presented by \citet{Grillo2015}.
In Figure \ref{fig:subhalo_distributions}, we also show the observed data from
\citet{Grillo2015}.
The color scheme is the same as in the previous plots.
Interestingly, the circular velocity function of the cluster members of
MACS~J1206 is more similar to that of MACS~J0416, a merging cluster, than to that
of RXC~J2248, a relaxed cluster.
In their work, \citet{Grillo2015} noticed how the number of massive
substructures ($v_c>100$ km s$^{-1}$) is underestimated in numerical simulations,
even when baryonic effects are included \citep{Munari2016}.
Similarly, we find that both in terms of radial and velocity distributions, the
results presented in this work are consistent with those by \citet{Grillo2015},
i.e.\ in tension with the predictions of numerical simulations, and partially at
odds with those by \citet{Natarajan2017}.
Only the circular velocity function of RXC~J2248 seems to be within the
intrinsic scatter of the simulated data; however, in the observed cluster the
number of sub-halos with a given circular velocity is still a factor of
approximately three larger than what found in simulations.
These differences can not be solved by invoking possible misidentifications of
cluster members, as our extensive spectroscopy and 12-band CLASH photometry 
lead to highly pure samples (i.e., very few false positives) and the vast
majority of the high-velocity cluster members are
confirmed spectroscopically \citep[see Fig. 4 in][]{Caminha2017a}.

\section{Conclusions} \label{sec:conclusions}
Thanks to the very high quality of the multi-wavelength data available, we
have been able to separate the collisional and collisionless mass components in
the galaxy clusters RXC~J2248.7$-$4431, MACS~J0416.1$-$2403, and
MACS~J1206.2$-$0847 at $z\approx0.4$.
Two of them, RXC~J2248 and MACS~J0416, are part of the HFF sample
\citep{Lotz2017} and all three have been observed with both the VIMOS
\citep[CLASH-VLT program;][]{Rosati2014} and the MUSE (archive observations)
instruments at the VLT.
Following the method presented by \citet{Bonamigo2017}, we have modeled the hot gas
mass by fitting the X-ray surface brightness from deep Chandra observations of
the clusters.
Then, we have introduced this mass term in the strong lensing analysis as a fixed component.

The main results of the work can be summarized as follows:
\begin{enumerate}
\item We have provided 2D models of the hot intracluster gas mass density of
these three clusters (updating that of MACS~J0416) that are consistent with a
well tested and independent approach \citep{Ettori2013} and that can be easily included in
different gravitational lensing softwares.
\item With the decoupling of the hot gas from the other cluster mass components,
we have improved previous strong-lensing models to describe more
accurately the different contributions to the total mass budget of the clusters.
Due to their different physical nature, the cluster hot gas and DM halo
components exhibit different properties, seen both in their surface mass density
maps and cumulative radial mass profiles.
\item The isolation of the diffuse DM component has allowed us to measure
with high accuracy the absence of a significant offset between the diffuse DM density peaks
and the positions of the BCGs, which is a test for models of self-interacting
dark matter.
\item By rescaling the radial profiles of the cluster projected mass with the
values of $R_{200\textrm{c}}$ and $M_{200\textrm{c}}$, we have shown that these clusters manifest
an almost homologous structure, despite their significantly different relaxation
status.
\item By exploiting the small statistical uncertainties on the cluster total
mass derived from our strong-lensing analysis, we have measured the hot gas
over total mass fraction throughout the core of the clusters with unprecedented
precision (less than $1\%$).
A remarkable advantage of the adopted approach is the possibility of
investigating spatially resolved maps of the gas fraction, in addition to the
traditional radial profiles.
\item Finally, we have confirmed the findings by \citet{Grillo2015} and
\citet{Munari2016} that current $N$-body simulations
under-predict the number of massive sub-halos ($v_c>90$ km s$^{-1}$) in
the cores of massive clusters.
This discrepancy is visible in all three clusters of our sample.
\end{enumerate}

In this paper we have shown the advantages of an accurate multi-wavelength study
of a well selected sample of clusters with high-quality data.
Being able to extend the sample to even more clusters and comparing with the
outcomes of cosmological simulations will allow to tackle some of the still open
questions about the nature of DM and the internal structure of galaxy clusters.
The importance of detailed and accurate studies of galaxy clusters is clear.
In this era of large all-sky surveys, the information gained from vast
samples of galaxy clusters rely on the accuracy of the adopted priors and
models.
Only by testing these assumptions on a smaller, well-understood sample it is
possible to push forward our knowledge of DM and the other components of the
Universe.

\acknowledgments
M.B. and C.G. acknowledge support by VILLUM FONDEN Young Investigator Programme
through grant no. 10123.
S.E. acknowledges the financial support from contracts ASI-INAF I/009/10/0,
NARO15 ASI-INAF I/037/12/0 and ASI 2015-046-R.0.
G.B.C. acknowledges funding from the ERC Consolidator Grant ID 681627-BUILDUP (P.I. Caputi).
G.B.C., P.R., A.M., M.A. and M.L. acknowledge financial support from PRIN-INAF
2014 1.05.01.94.02.
AM acknowledges funding from the INAF PRIN-SKA 2017 program 1.05.01.88.04.
Simulated data postprocessing and storage has been done on the CINECA facility
PICO thanks to che Iscra C grant GALPP\_3.

\appendix
\section{Alternative hot gas mass models}\label{sec:xray_models}
When fitting the X-ray surface brightness, we model the cluster hot gas
mass density distributions with either two or three components (three and four for MACS~J0416),
with both spherical and elliptical symmetry.
The selection of the final best-fitting model is done by considering the Akaike information criterion
\citep[AIC,][]{Akaike1974} and the Bayesian Information Criterion
\citep[BIC,][]{Schwarz1978}.
As here we are mainly interested in the projected mass profiles of the clusters, we compare
our results with independent measurements of the cumulative projected hot gas
mass profiles of RXC~J2248, MACS~J0416, and MACS~J1206.
\begin{figure}
	\centering
	\includegraphics[width=\textwidth]{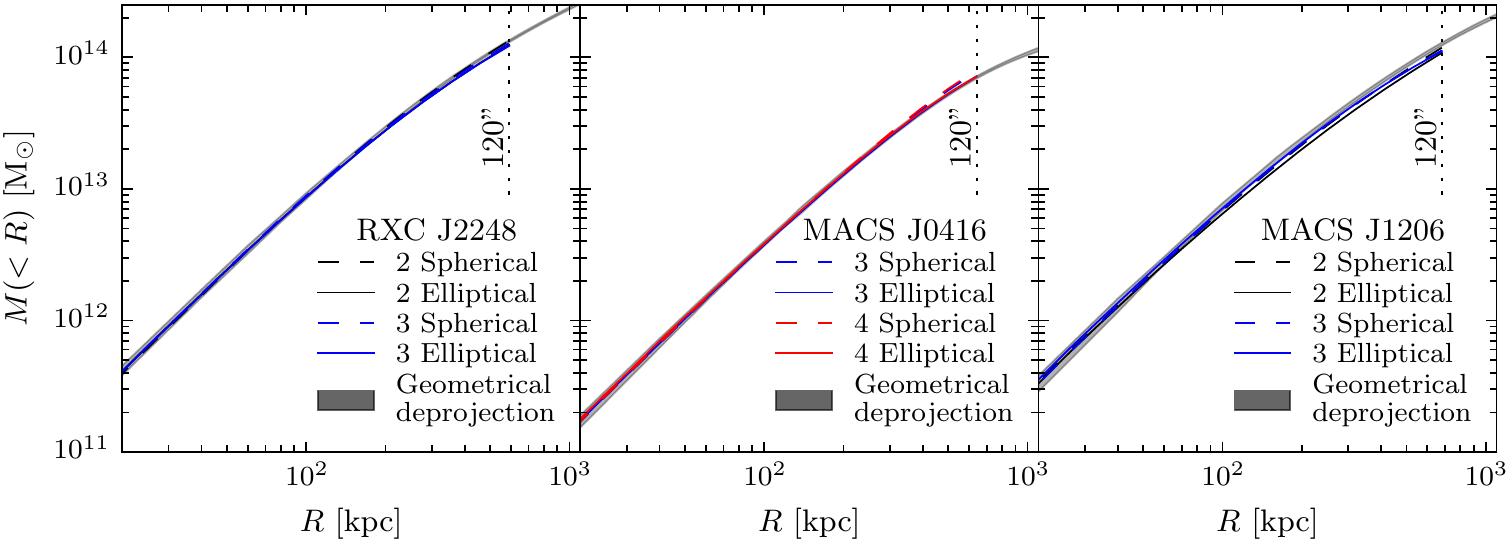}
    \caption{Cumulative projected hot gas mass profiles of RXC~J2248, MACS~J0416
    and MACS~J1206, respectively, on the left, center and right panel. Dashed and
    solid lines show, the spherical- and elliptical-component models, respectively.
    Black, blue and red lines refer to models with two, three and four components.
    The gray areas show the $1$-sigma confidence regions of the mass profile
    obtained through a geometrical deprojection.}
	\label{fig:mass_profile_gas}
\end{figure}
In Fig. \ref{fig:mass_profile_gas}, the gray areas show the $1$-sigma confidence
regions of these mass profiles, while the dashed and solid lines show,
respectively, the spherical- and elliptical-component models tested in this
work \citep[based on][]{Bonamigo2017}.
Different colors represent models with different numbers of components, as
indicated on the legends.
The model used as comparison has been recovered through the geometrical
deprojection \citep[see, e.g.,][and references therein]{Ettori2013} of the
azimuthally averaged surface brightness profile that considers the entire X-ray
emission of the cluster. 
The striking agreement with this different and independent method corroborates
the accuracy of our measurements, especially in the regions of interest in the
lensing analysis, i.e. $R<400$~kpc.

\section{Effect of the intrinsic variance of the $\chi^2$ distribution}\label{sec:chi2distribution}
As any other distribution, the $\chi^2$ distribution has an intrinsic
variance that is a function of the model degrees of freedom.
Because of this, even a perfect model is not guaranteed to have a value of the $\chi^2$
equal to the number of the degrees of freedom \citep{andrae2010}.
Therefore, when we require the best-fit model to have a $\chi^2$ value
approximately equal to the number of degrees of freedom, we introduce a
systematic error, that propagates into the uncertainties on the mass-component
parameter distributions and on the mass and density radial profiles.

To better quantify this effect, we consider the value of the $0.13$
percentile ($99.7\%$ confidence level) of the $\chi^2$ distribution for our
best-fit strong-lensing model of RXC~J2248.
This gives a conservative model that can be considered as an upper limit on the
errors of the mass models presented in this paper.
Given the $59$ degrees of freedom for RXC~J2248, the $0.13$-percentile of the
corresponding $\chi^2$ distribution is $31.7$.
Therefore, to investigate this more conservative model, we increase the uncertainty
on the image positions to $0''.66$, in order to get a minimum chi-square value
of approximately 32, while in the main-text model the error is $0''.48$.

\begin{figure}
	\centering
	\includegraphics[width=\textwidth]{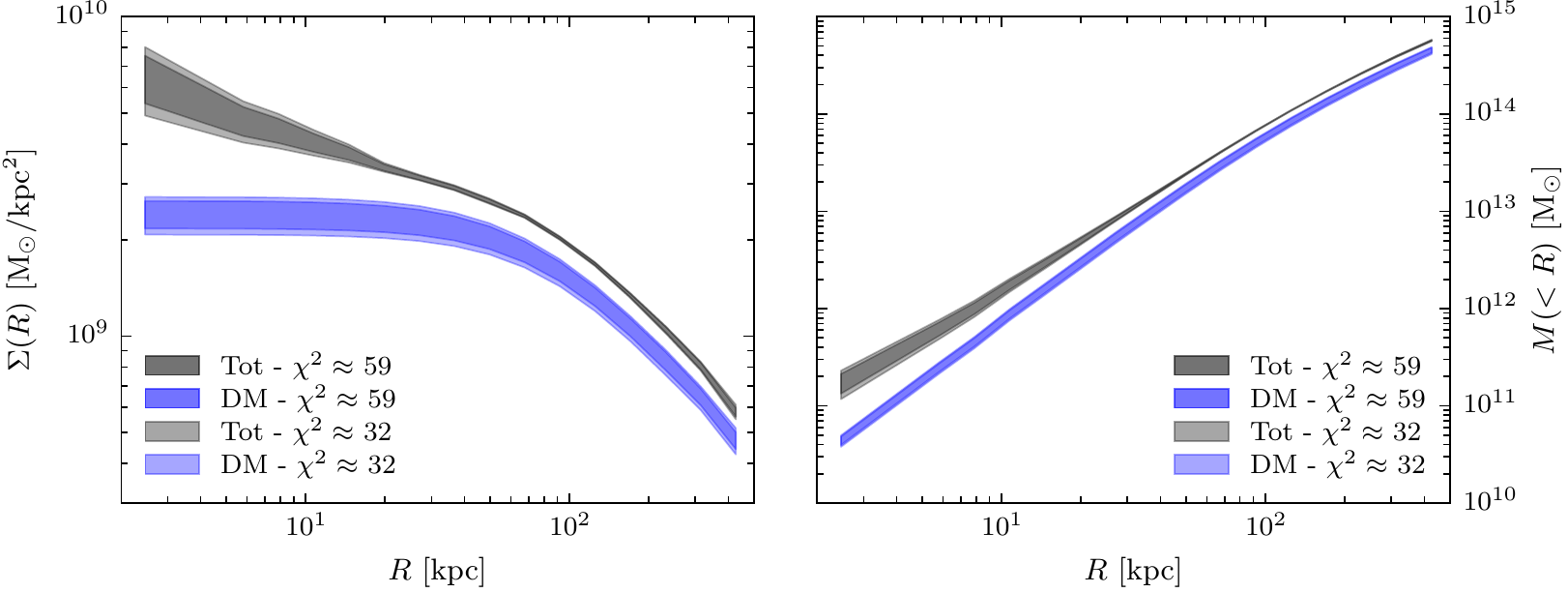}
	\caption{Radial profiles of the surface mass density (left panel) and
	cumulative projected mass (right panel) for the total (black) and
	dark-matter component (blue) of RXC~J2248. The areas show values at the
	3$\sigma$ confidence level. Darker and lighter refer to values of the
	minimum $\chi^2$ of approximately $59$ (reference model in this work) and
	$32$, respectively.}
	\label{fig:mass_decomposition_chi2}
\end{figure}
We then use the same procedure presented above to compute the surface
mass density and cumulative projected mass radial profiles and show them,
respectively, on the left and right panel of Figure \ref{fig:mass_decomposition_chi2}.
The filled areas represent the $3\sigma$ uncertainties on the radial profiles of the
main-text (darker colors) and conservative (lighter colors) models, respectively.
Black and blue areas show the total and DM-only component.

We remark that the $3\sigma$ confidence level values of the profiles
shown in Figure \ref{fig:mass_decomposition_chi2} for the two different models are
very similar. The results of this test thus suggest that the variance of the
$\chi^2$ distribution does not affect significantly the final errors on the
values of the parameters and of the derived quantities of a model.

\section{On the homologous projected profiles}\label{sec:nfw_profiles}
As noted in Section \ref{sec:discussion}, the rescaled surface density and
cumulative projected mass profiles of the three clusters are very similar,
suggesting the existence of a homologous mass profile.
Indeed, DM-only numerical simulations have shown that the averaged mass profile
of virialized halos is well described by a universal profile, the so-called
Navarro-Frenk-White profile \citep[NFW;][]{Navarro1997}.
We use a NFW profile to fit, separately, both the surface mass density and the
cumulative projected mass profiles estimated in our strong lensing analyses.
\begin{figure}
	\centering
	\includegraphics[width=\textwidth]{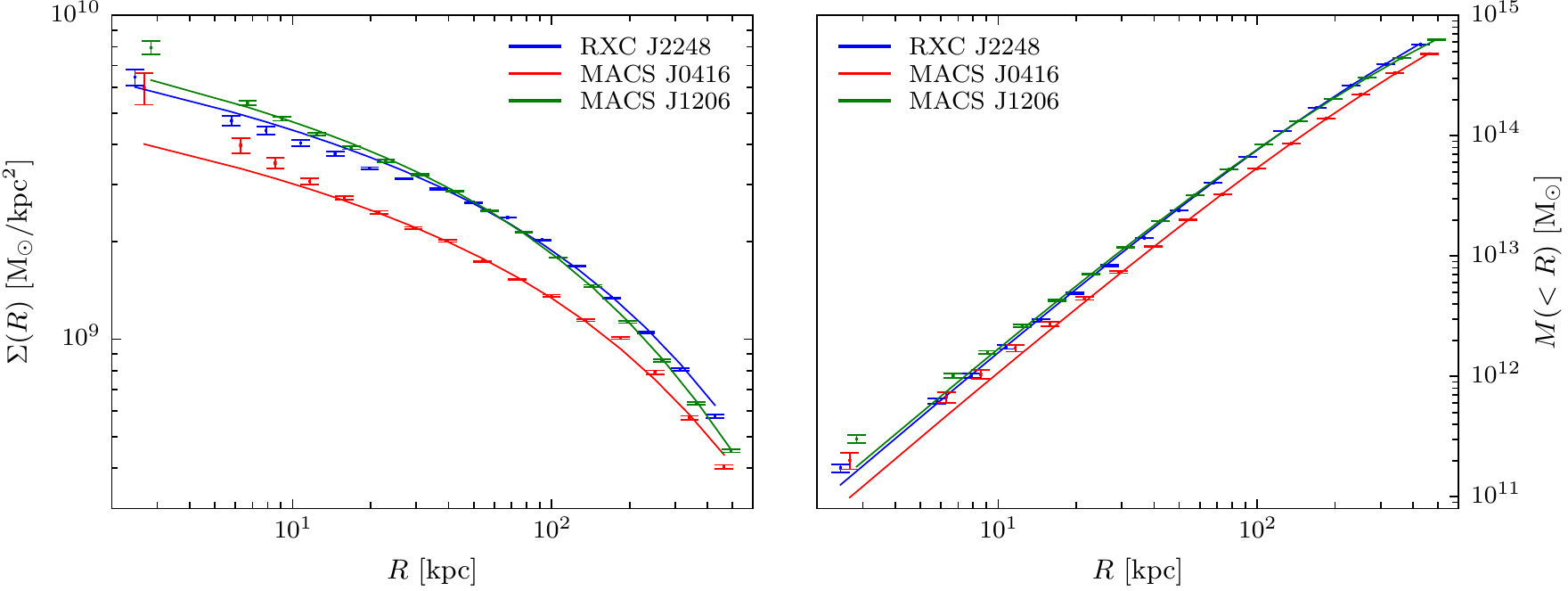}
    \caption{Radial profiles of the total surface mass density (left panel) and
        cumulative projected mass (right panel) of RXC~J2248
	    (blue), MACS~J0416 (red) and MACS~J1206 (green). Data points with
	    error bars show the fitted data, measured from the strong lensing
	    analysis. Solid lines represent the best-fit NFW profiles to the data.}
	\label{fig:nfw_fits}
\end{figure}
The best-fitting profiles obtained from the optimized values are shown in Fig. \ref{fig:nfw_fits}
as solid lines.
Points with error bars show the mean and standard deviation of the measured
surface mass density (left panel) and cumulative projected mass (right panel)
that have been fitted.
With the exception of the innermost regions, dominated by the BCGs, a NFW
profile provides a good fit to the data, as shown in \citet{Umetsu2014} for the
CLASH cluster sample.
We remark though that the data points are obtained from the combination of multiple components, described
by different (cored) isothermal mass profiles, which are favored over the NFW profiles by the strong
lensing analyses, as shown in \citet{Grillo2015} and \citet{Caminha2017b} for MACS~J0416 and MACS~J1206, respectively.
This suggests that other one-component models, with a varying radial slope, might also provide good fits to the reconstructed profiles of the clusters in their cores, when they are considered in projection. We believe that the best way to distinguish among the different models is to use the chi-square statistics directly on the difference between the observed and model-predicted positions of the multiple images and that a good fit on the reconstructed projected quantities of a complex astrophysical object might not be enough to state the definite success of a particular model.

\bibliography{library}

\listofchanges

\end{document}